\documentclass[conference]{IEEEtran}
\IEEEoverridecommandlockouts

\usepackage{cite}
\usepackage{amsmath,amssymb,amsfonts}
\usepackage{graphicx}
\usepackage{textcomp}
\usepackage{xcolor}

\usepackage{algorithm, algpseudocode}
\usepackage{xcolor}
\usepackage{tabularx, bm}
\usepackage{caption, subcaption}

\usepackage{colortbl}

\usepackage{listings}

\usepackage{tcolorbox}

\usepackage{booktabs}
\usepackage{pifont}
\usepackage{listings}
\usepackage{caption}
\usepackage{subcaption}
\usepackage{hyperref}
\usepackage[belowskip=-2pt,aboveskip=6pt]{caption}

\newcommand{\lambdab}{\bm{\lambda}}
\newcommand{\mub}{\bm{\mu}} 
\newcommand{\xb}{\mathbf{x}}
\newcommand{\zb}{\mathbf{z}}
\newcommand{\cb}{\mathbf{c}}
\newcommand{\db}{\mathbf{d}}
\newcommand{\bb}{\mathbf{b}}
\newcommand{\Ab}{\mathbf{A}}
\newcommand{\Bb}{\mathbf{B}}

\newcommand{\Qb}{\mathbf{Q}}
\newcommand{\Ib}{\mathbf{I}}
\newcommand{\xbl}{\underline{\mathbf{x}}}
\newcommand{\xbu}{\overline{\mathbf{x}}}
\newcommand{\zbl}{\underline{\mathbf{z}}}
\newcommand{\zbu}{\overline{\mathbf{z}}}
\newcommand{\Nc}{\mathcal{N}}
\newcommand{\Gc}{\mathcal{G}}
\newcommand{\Pc}{\mathcal{P}}
\newcommand{\Dc}{\mathcal{D}}
\newcommand{\Yc}{\mathcal{Y}}
\newcommand{\Ec}{\mathcal{E}}
\newcommand{\Lc}{\mathcal{L}}
\newcommand{\Ic}{\mathcal{I}}
\newcommand{\pg}{p^{\text{g}}}
\newcommand{\qg}{q^{\text{g}}}
\newcommand{\pb}{p^{\text{b}}}
\newcommand{\qb}{q^{\text{b}}}
\newcommand{\pd}{p^{\text{d}}}
\newcommand{\qd}{q^{\text{d}}}
\newcommand{\pgl}{\underline{p}^{\text{g}}}
\newcommand{\pgu}{\overline{p}^{\text{g}}}
\newcommand{\qgl}{\underline{q}^{\text{g}}}
\newcommand{\qgu}{\overline{q}^{\text{g}}}
\newcommand{\pl}{\underline{p}}
\newcommand{\pu}{\overline{p}}
\newcommand{\ql}{\underline{q}}
\newcommand{\qu}{\overline{q}}
\newcommand{\wl}{\underline{w}}
\newcommand{\wu}{\overline{w}}

\newcommand{\thetal}{\underline{\theta}}
\newcommand{\thetau}{\overline{\theta}}
\newcommand{\gsh}{g^{\text{sh}}}
\newcommand{\bsh}{b^{\text{sh}}}
\newcommand{\gs}{g^{\text{s}}}
\newcommand{\bs}{b^{\text{s}}}
\newcommand{\Mp}{M^{\text{p}}}
\newcommand{\Mq}{M^{\text{q}}}

\definecolor{codegreen}{rgb}{0,0.6,0}
\definecolor{codegray}{rgb}{0.5,0.5,0.5}
\definecolor{codepurple}{rgb}{0.58,0,0.82}
\definecolor{backcolour}{rgb}{0.95,0.95,0.92}

\lstdefinestyle{mystyle}{
    backgroundcolor=\color{backcolour},   
    commentstyle=\color{codegreen},
    keywordstyle=\color{magenta},
    numberstyle=\tiny\color{codegray},
    stringstyle=\color{codepurple},
    basicstyle=\ttfamily\footnotesize,
    breakatwhitespace=false,         
    breaklines=true,                 
    captionpos=b,                    
    keepspaces=true,                 
    numbers=left,                    
    numbersep=5pt,                  
    showspaces=false,                
    showstringspaces=false,
    showtabs=false,                  
    tabsize=2,
    xleftmargin=1.2em 
}

\def\BibTeX{{\rm B\kern-.05em{\sc i\kern-.025em b}\kern-.08em
    T\kern-.1667em\lower.7ex\hbox{E}\kern-.125emX}}
\begin{document}

\title{A GPU-Accelerated Distributed Algorithm for Optimal Power Flow in Distribution Systems\\
}

\author{\IEEEauthorblockN{Minseok Ryu and Geunyeong Byeon}
\IEEEauthorblockA{\textit{School of Computing and Augmented Intelligence} \\
\textit{Arizona State University}\\
Tempe, AZ, USA \\
\{mryu2, geunyeong.byeon\}@asu.edu}
\and
\IEEEauthorblockN{Kibaek Kim}
\IEEEauthorblockA{\textit{Mathematics and Computer Science Division} \\
\textit{Argonne National Laboratory}\\
Lemont, IL, USA \\
kimk@anl.gov}
}

\maketitle

\begin{abstract}
We propose a GPU-accelerated distributed optimization algorithm for controlling multi-phase optimal power flow in active distribution systems with dynamically changing topologies. To handle varying network configurations and enable adaptable decomposition, we advocate a componentwise decomposition strategy. However, this approach can lead to a prolonged computation time mainly due to the excessive iterations required for achieving consensus among a large number of fine-grained components. To overcome this, we introduce a technique that segregates equality constraints from inequality constraints, enabling GPU parallelism to reduce per-iteration time by orders of magnitude, thereby significantly accelerating the overall computation. Numerical experiments on IEEE test systems ranging from 13 to 8500 buses demonstrate the superior scalability of the proposed approach compared to its CPU-based counterparts. 
\end{abstract}


\section{Introduction} \label{sec:intro} 
The modern evolution of electric power distribution systems, exemplified by the rising integration of renewable and distributed energy resources such as solar photovoltaic, energy storage systems, and electric vehicles, is imposing a greater need for advanced operational and planning strategies based on distributed Optimal Power Flow (OPF).
The significance of distributed OPF solution within the distribution systems is expanding due to its inherent benefits in terms of scalability, adaptability, privacy, and robustness in comparison to centralized optimization \cite{patari2021distributed}.  

Distributed OPF problem (e.g., \cite{kim1997coarse, baldick1999fast,biskas2006decentralised, Kim_2000, sun2013fully, peng2014distributed, Erseghe_2014,  mhanna2018adaptive, muhlpfordt2021distributed, ryu2021privacy, alkhraijah2022assessing, ryu2023differentially, Inaolaji_2023, pinto2020distributed}) is often formulated as the following consensus optimization model:
\begin{subequations}
\label{DO_model}
\begin{align}
\min \ &  f (\xb) \label{DO_model_obj} \\
\mbox{s.t.} \ & \xb_s \in \mathcal{X}_s, \ \forall s \in [S], \label{DO_model_const} \\
& \Bb_s \xb = \xb_s, \ \forall s \in [S]. \label{DO_model_consensus}
\end{align}
\end{subequations}
Here, $\xb$ represents global OPF variables for the entire network, 
$f$ is the objective function (e.g., total operational cost), and
$\mathcal{X}_s$ is a feasible region of $\xb_s$, a copy of some variables in $\xb$ representing the local OPF variables defined for each subsystem $s \in [S]:=\{1, \ldots, S\}$ of the entire power network.
Eq. \eqref{DO_model_consensus} refers to consensus constraints that establish interconnections among subsystems within the network through the matrix $\Bb_s$ whose elements are $0$ or $1$, thereby guaranteeing the feasibility of the original distributed OPF solutions acquired. 

Various distributed optimization algorithms have been proposed for solving \eqref{DO_model}, contingent upon the chosen network decomposition strategy, resulting in different $S$, and the specific modeling of OPF, resulting in different $\mathcal{X}_s$. The widely used algorithm is the alternating direction method of multipliers (ADMM) which is an iterative algorithm that solves the augmented Lagrangian formulation of \eqref{DO_model}, derived by penalizing \eqref{DO_model_consensus} in the objective function. At each iteration of ADMM, it is possible to solve subproblems in parallel, each necessitating the utilization of optimization solvers running on CPU. 
 
Among the various network decomposition strategies \cite{kim1997coarse, baldick1999fast,biskas2006decentralised}, the component-wise decomposition \cite{sun2013fully, mhanna2018adaptive}, which partitions the power network into individual components (e.g., generators, buses, lines, and transformers), enables fully decentralized and distributed OPF computations at the component level. This approach is particularly advantageous for adaptive control systems in dynamically changing network topologies, as it offers flexibility in incorporating or omitting components from control regions. 
With this approach, the distributed OPF model \eqref{DO_model} typically involves large $S$ (i.e., the number of components) while the feasible region $\mathcal{X}_s$ for each component $s \in [S]$ is defined by relatively few constraints.
Although the OPF at each component can be computed by solving smaller subproblem in parallel \eqref{DO_model_const}, achieving consensus (i.e., \eqref{DO_model_consensus}) among a large number $S$ of components is often more time-consuming, resulting in high overall computation time to obtain a feasible OPF solution for the entire network.

To enhance the computational efficiency, this paper proposes a GPU-accelerated ADMM for solving the distributed OPF model \eqref{DO_model}, where $\mathcal{X}_s$ is modeled using a linear approximation \cite{byeon2022linear}, and a component-wise decomposition strategy is employed to accommodate flexible network configurations in electric power distribution systems.
Specifically, we 
isolate all bound constraints from $\mathcal{X}_s$ and incorporate them into the global update step of ADMM. 
By limiting the local subproblems to linear equality constraints, we replace the need for optimization solvers with simple matrix operations,  which GPUs can handle with exceptional performance. 

\subsection{Related work and contributions}
Various distributed OPF models and algorithms (e.g., \cite{kim1997coarse, baldick1999fast,biskas2006decentralised, Kim_2000, sun2013fully, peng2014distributed, Erseghe_2014,  mhanna2018adaptive, muhlpfordt2021distributed, ryu2021privacy, alkhraijah2022assessing, ryu2023differentially, Inaolaji_2023, pinto2020distributed}) have been proposed in the literature.
While most literature focuses on distributed OPF in transmission systems where a single phase OPF is considered, a few exception exists for distributed multi-phase OPF in distribution systems.
For example, the authors in \cite{pinto2020distributed} solves a component-wise distributed multi-phase OPF model using ADMM.
However, the algorithm has experienced prolonged computation time to convergence, even for a small-size 33-bus system instance, mainly due to the expensive (convex) subproblem optimization processes that requires optimization solvers. 
In this work, we adopt our recent development \cite{byeon2022linear} on the linear programming approximation for a multi-phase OPF with delta connections to formulate the component-wise distributed model, and develop a solver-free, GPU-accelerated ADMM algorithm.
To the best of our knowledge, this is the first work that leverages GPU to accelerate solver-free ADMM for solving component-wise distributed multi-phase OPF model that scales to large systems, such as the IEEE 8500-bus system. 

Recently, several GPU-accelerated optimization algorithms \cite{kim2021leveraging,pacaud2024accelerating,pacaud2024parallel, schubiger2020gpu} and their applications for solving OPF problem \cite{rakai2014gpu, zhou2017gpu, kim2022accelerated, shin2024accelerating} have been proposed.
In particular, a GPU has been leveraged to accelerate \emph{solver-free} ADMM in \cite{schubiger2020gpu} which is built upon the popular OSQP \cite{osqp} for solving a general quadratic programming with linear constraints $\Ab \xb=\zb$ and bound constraints $\zb \in [\zbl, \zbu]  $. 
This algorithm is solver-free primarily because a unique closed-form solution to the subproblem exists even when $\Ab$ is not a full row rank matrix. This is achieved by introducing additional variables $(\tilde{\xb}, \tilde{\zb})$, which are copies of the original variables $(\xb, \zb)$, ensuring that the resulting KKT matrix is invertible.
However, employing the GPU-accelerated solver-free ADMM in \cite{schubiger2020gpu} for solving the component-wise distributed OPF model is not direct as it requires decomposing the matrix $\Ab$ into smaller submatrices $\Ab_s$ for each subsystem.
In this work, 
instead of introducing additional variables to make the KKT matrix invertible, we leverage the fine-grained decomposition that produces small $\mathbf{A}_s$ matrices, allowing us to directly apply row reduction operations to ensure that $\mathbf{A}_s$ has full row rank, thus guaranteeing a closed-form unique solution for each subproblem.

Our contributions are summarized as follows:
\begin{enumerate}
  \item development of a solver-free ADMM algorithm for solving a distributed multi-phase linearized OPF in electric power distribution systems,     
  \item implementation of the algorithm that can efficiently run on NVIDIA GPU architecture, as well as CPUs in parallel; and  
  \item demonstration of the greater computational performance of the proposed GPU-accelerated algorithm compared to the multi-CPU counterparts.
\end{enumerate}

\subsection{Paper structure}
The remainder of this paper is organized as follows. 
In Section \ref{sec:math_form}, we describe the mathematical formulation for the distributed multi-phase OPF model for electric power distribution systems (more details in \cite{byeon2022linear}).
Then we present the proposed GPU-accelerated solver-free ADMM in Section \ref{sec:proposed}.
Finally, in Section \ref{sec:experiments}, we numerically showcase the superiority of our algorithm using IEEE test systems ranging from 13 to 8500 buses demonstrating the superior scalability compared to its CPU-based counterparts. 

\begin{table} 
\centering
\caption{Nomenclature}  
\begin{tabularx}{0.5\textwidth}{ll}      
\hline
\multicolumn{2}{c}{Sets} \\ \hline
  $\Nc$, $\Gc$, $\Ec$, $\mathcal L$ & a set of buses, generators, lines, and loads\\  
  $\Lc_i$, $\Yc_i$, $\Dc_i$ & a set of loads, wye loads, and delta loads at bus $i \in \Nc$ \\
  $\Pc_c \subseteq \{1,2,3\}$ & a set of phases of component $c \in \Nc \cup \Gc \cup \Ec \cup \Lc$ \\
  \hline 
  \multicolumn{2}{c}{Parameters} \\ \hline
  $\pgl_{k\phi} , \pgu_{k\phi}, \qgl_{k\phi} , \qgu_{k\phi}$ & bounds on the $\phi$-phase power generation of $k \in \Gc$.  \\
  $\wl_{i\phi}, \wu_{i\phi}$ & bounds on the squared magnitude of voltage  \\ & at $i \in \Nc$ and $\phi \in \Pc_i$. \\
  $\gsh_{i \phi}, \bsh_{i \phi}$ & shunt conductance and shunt susceptance \\ & 
  at $i \in \Nc$ and $\phi \in \Pc_i$. \\  
  $r_{e \phi \phi'}, x_{e \phi \phi'}$ & resistance and reactance at $e \in \Ec$ and $\phi, \phi' \in \Pc_{e}$. \\
  $\thetal_{e}, \thetau_{e}$ & bounds on the angle difference at line $e \in \Lc$.\\
  $g^{\text{s}}_{e i j \phi}, b^{\text{s}}_{e ij \phi }$ & shunt conductance and shunt susceptance \\ & at from-bus $i$ of $e \in \Ec$ on phase $\phi \in \Pc_{e}$. \\
  $\tau_{e \phi}$ & tap ratio at $e \in \Ec$ and $\phi \in \Pc_{e}$. \\       
\hline
\multicolumn{2}{c}{Variables} \\ \hline
$\pg_{k\phi}, \qg_{k\phi}$ & $\phi$-phase power generated by $k \in \Gc$.\\
$w_{i\phi} $ & $\phi$-phase voltage magnitude squared at bus $i \in \Nc$. \\
$\pd_{l\phi}, \qd_{l\phi}$ & $\phi$-phase power consumption of load $l \in \Lc$.  \\
$\pb_{l\phi},  \qb_{l\phi}$ & $\phi$-phase power withdrawn by load $l \in \Lc$ from the \\
& bus it is attached. \\
$p_{eij \phi}, q_{e ij \phi}$ & $\phi$-phase power flow on line $e_{ij} \in \Ec$. \\ \hline
\end{tabularx}  
\label{Table:Nomenclature}
\end{table}

\section{Mathematical Formulations} \label{sec:math_form}
In this section we present both centralized and distributed multi-phase linearized OPF model with delta connection, proposed in \cite{byeon2022linear}.
\subsection{Centralized multi-phase OPF model}
Based on the notations in Table \ref{Table:Nomenclature}, we present a centralized multi-phase OPF model (see \cite{byeon2022linear} for more details).
\subsubsection{Operational bound constraints} OPF needs to ensure that power generation, voltage magnitude, and power flow, respectively, are within certain bounds:
\begin{subequations}
\label{operational}
\begin{align}
& \pg_{k\phi} \in [\pgl_{k\phi},\pgu_{k\phi}], \qg_{k\phi} \in [\qgl_{k\phi},\qgu_{k\phi}], \ \forall k \in \Gc, \phi \in \Pc_k, \\
& w_{i\phi} \in [\wl_{i\phi}, \wu_{i\phi}], \ \forall i \in \Nc,  \phi \in \Pc_i, \\
& p_{eij\phi},p_{eji\phi} \in [\pl_{eij\phi},\pu_{eij\phi}], \ \forall (e,i,j) \in \Ec, \phi \in \Pc_e, \\
& q_{eij\phi},q_{eji\phi} \in [\ql_{eij\phi},\qu_{eij\phi}], \ \forall (e,i,j) \in \Ec, \phi \in \Pc_e.
\end{align}
\end{subequations}

\subsubsection{Power balance equations}
For every bus $i \in \Nc$ and phase $\phi \in \Pc_i$, the power balance equations are given by
\begin{subequations}
\label{balance_eqns}
\begin{align}
& \sum_{ (e, i,j) \in \Ec_i } p_{e i j  \phi} + \sum_{l \in \Lc_i} \pb_{l \phi} + \gsh_{i \phi} w_{i \phi} = \sum_{ k \in \Gc_i } \pg_{k \phi}, \\
& \sum_{ (e, i,j) \in \Ec_i } q_{e i j \phi} + \sum_{l \in \Lc_i} \qb_{l \phi}  - \bsh_{i \phi} w_{i \phi} = \sum_{ k \in \Gc_i } \qg_{k \phi},  
\end{align}  
\end{subequations}
which ensure power balance at every bus in the network.
Note that $\pb_{l\phi}$ and $\qb_{l\phi}$ represent real and reactive power withdrawals on phase $\phi$ by a load $l$ from the bus to which it is attached. 

\subsubsection{Voltage dependent load model}
For every bus $i \in \Nc$, the voltage dependent load model is given by  
\begin{subequations}
\label{volt_load_model}  
\begin{align}
& \pd_{l\phi} = \frac{a_{l\phi} \alpha_{l\phi}}{2} (\widehat{w}_{l\phi}-1)+ a_{l\phi}, \ \forall l \in \Lc_i, \forall \phi \in \Pc_i, \label{VDLM-1}\\
& \qd_{l\phi} = \frac{b_{l\phi} \beta_{l\phi}}{2} (\widehat{w}_{l\phi}-1) + b_{l\phi},  \   \forall l \in \Lc_i, \forall \phi \in \Pc_i, \label{VDLM-2}\\
&  \widehat{w}_{l \phi} = w_{i\phi}, \  \forall l \in \Yc_i, \forall \phi \in   \Pc_i, \label{VDLM-3}\\
&  \widehat{w}_{l \phi} = 3 w_{i\phi}, \ \forall l \in \Dc_i, \forall \phi \in \Pc_i, \label{VDLM-4}
\end{align}
where, for fixed $l \in \Lc_i$ and $\phi \in \Pc_i$,  
$\pd_{l\phi}$ and $\qd_{l\phi}$ are real and reactive power consumption,  
$\alpha_{l\phi}$ and $\beta_{l\phi}$ are nonnegative scalars given as inputs that determines the load type, 
$a_{l\phi}$ and $b_{l\phi}$ are also given input data determined by reference values for real power load, reactive power load, and voltage magnitude applied to the load, 
$\widehat{w}_{l\phi}$ in \eqref{VDLM-1} and \eqref{VDLM-2} are the squared magnitude of the voltage applied to the load.

For the wye connected load $l \in \Yc$ on phase $\phi \in \Pc_l$, the relationship between $\pb$ and $\pd$ is:
\begin{align}
\pb_{l \phi} = \pd_{l \phi}, \ \ \qb_{l \phi} = \qd_{l \phi}. \label{VDLM-5}
\end{align}
For the delta connected load, $l \in \Dc$, the relationship between $\pb$ and $\pd$ is given by 
\begin{align} 
& \sum_{\phi \in \Pc_l} \pb_{l\phi} - \pd_{l\phi} = \sum_{\phi \in \Pc_l} \qb_{l\phi} - \qd_{l\phi} = 0,  \label{VDLM-6}\\ 
&  \frac{3}{2} \pb_{l2} - \frac{\sqrt{3}}{2} \qb_{l2}  = {\pd_{l2} + \frac{1}{2} \pd_{l1} - \frac{\sqrt{3}}{2} \qd_{l1}},  \label{VDLM-7}\\
& \frac{\sqrt{3}}{2} \pb_{l2} + \frac{3}{2} \qb_{l2} = {\frac{\sqrt{3}}{2} \pd_{l1} +\frac{1}{2} \qd_{l1} + \qd_{l2}},  \label{VDLM-8}\\
& \sqrt{3} \qb_{l2} + \frac{3}{2} \pb_{l3}-\frac{\sqrt{3}}{2} q^{b}_{l3} = {\frac{1}{2} \pd_{l1} + \frac{\sqrt{3}}{2} \qd_{l1} + \pd_{l3}},  \label{VDLM-9} \\
& - \sqrt{3}\pb_{l2} + \frac{\sqrt{3}}{2} \pb_{l3} + \frac{3}{2} q^{b}_{l3}   = {- \frac{\sqrt{3}}{2} \pd_{l1} + \frac{1}{2} \qd_{l1}  + \qd_{l3}}.  \label{VDLM-10} 
\end{align}
\end{subequations}

\subsubsection{Linearized power flow equations}
For every line $(e,i,j) \in \Ec$ and phase $\phi \in \Pc_e$, the linearized power flow equations are given by
\begin{subequations}
\label{power_flow_eqns}  
\begin{align} 
& p_{e ij \phi} + p_{e ji \phi} = \gs_{e i j \phi} w_{i \phi} + \gs_{e ji \phi} w_{j \phi}, \label{powerloss-1} \\
& q_{e ij \phi} + q_{e ji \phi} = -\bs_{e ij \phi}  w_{i \phi} -\bs_{e ji \phi} w_{j \phi}, \label{powerloss-2} \\ 
& w_{i \phi} = \tau_{e \phi} w_{j \phi} - \sum_{\psi \in \Pc_{e}} \Mp_{e \phi \psi} (p_{e ij \psi}-\gs_{e ij \psi} w_{i \psi}) \nonumber \\
& \ \ \ \ \ \ \ \ \ \ \ \  - \sum_{\psi \in \Pc_{e}} \Mq_{e \phi \psi} (q_{e ij \psi}+\bs_{e ij \psi} w_{i \psi}), \label{voltage_mag_diff}   
\end{align}
\end{subequations}
where $\Mp_{e \phi \psi}$ and $\Mq_{e \phi \psi}$ are $(\phi, \psi)$-th element of $\Mp_{e}$ and $\Mq_{e}$, respectively, which are defined as

\begin{align*} 
  & \Mp_{e} = 
  \begin{bmatrix}
    -2r_{\ell 11}  & r_{\ell 12} - \sqrt{3}x_{\ell 12} & r_{\ell 13} + \sqrt{3}x_{\ell 13} \\
    r_{\ell 21} + \sqrt{3}x_{\ell 21} & -2r_{\ell 22} & r_{\ell 23} - \sqrt{3}x_{\ell 23} \\ 
    r_{\ell 31} - \sqrt{3} x_{\ell 31} & r_{\ell 32} + \sqrt{3}x_{\ell 32} & -2r_{\ell 33}
  \end{bmatrix} \\
  & \Mq_{e} = 
  \begin{bmatrix}
    -2x_{\ell 11}  & x_{\ell 12} + \sqrt{3}r_{\ell 12} & x_{\ell 13} - \sqrt{3}r_{\ell 13} \\
    x_{\ell 21} - \sqrt{3} r_{\ell 21} & -2x_{\ell 22} & x_{\ell 23} + \sqrt{3}r_{\ell 23} \\ 
    x_{\ell 31} + \sqrt{3} r_{\ell 31} & x_{\ell 32} - \sqrt{3}r_{\ell 32} & -2x_{\ell 33}
  \end{bmatrix} 
\end{align*}

\subsubsection{Multi-phase OPF model}
To summarize, we have 
\begin{subequations}
\label{lindist3flow}  
\begin{align} 
\min \ &  \sum_{k \in \Gc} \sum_{\phi \in \Pc_k} \pg_{k \phi} \label{lindist3flow_obj}\\
\mbox{s.t.} \ 
& \eqref{operational}, \eqref{balance_eqns}, \eqref{volt_load_model}, \eqref{power_flow_eqns},  
\end{align}
\end{subequations}
which can be represented as the following abstract LP form:
\begin{subequations}
\label{LP_model}
\begin{align}
\min \ & \cb^{\top} \xb \label{LP_model_obj} \\
\mbox{s.t.} \ & \Ab \xb = \bb, \label{LP_model_eqn}\\
\ & \xbl \leq \xb \leq \xbu, \label{LP_model_bnd}
\end{align}
where
\begin{align}
\xb = 
\begin{bmatrix}
  & \{\pg_{k\phi}, \qg_{k\phi}\}_{k\in\Gc, \phi \in \Pc_k}   \\
  & \{ w_{i\phi}\}_{i\in\Nc, \phi \in \Pc_i} \\
  & \{\pb_{l\phi}, \qb_{l\phi}, \pd_{l\phi}, \qd_{l\phi} \}_{l \in \Lc_i,  \phi \in \Pc_l } \\
  & \{ p_{eij\phi}, q_{eij\phi}, p_{eji\phi}, q_{eji\phi} \}_{(e,i,j) \in \Ec, \phi \in \Pc_e}
\end{bmatrix}, \nonumber
\end{align}
\end{subequations}
\eqref{LP_model_obj} corresponds to \eqref{lindist3flow_obj},
\eqref{LP_model_eqn} corresponds to \eqref{balance_eqns}, \eqref{volt_load_model}, \eqref{power_flow_eqns}, 
and
\eqref{LP_model_bnd} corresponds to \eqref{operational}.

\subsection{Distributed multi-phase OPF model} \label{sec:dist_opf}
With the componentwise decomposition strategy, the model \eqref{LP_model} can be rewritten as 
\begin{subequations}
\label{LP_model_decomposed}
\begin{align}
\min \ & \cb^{\top} \xb \\
\mbox{s.t.} \ 
& \Ab_s \xb_s = \bb_s, \ \ \xb_s \in [\xbl_s, \xbu_s], \ \forall s \in [S],  \\
& \Bb_s \xb = \xb_s, \ \forall s \in [S],
\end{align}
\end{subequations}
where, for each component $s \in [S]$, $\Ab_s \in \mathbb{R}^{m_s \times n_s}$, $\bb_s \in \mathbb{R}^{m_s}$, $\xbl_s, \xbu_s \in \mathbb{R}^{n_s}$ are given input data, and $\Bb_s \in \mathbb{R}^{n_s \times n}$ is a given 0-1 matrix wherein each row sums to $1$, while each column sums to either $0$ or $1$.
This matrix serves as a mapping operator that maps global variables $\xb \in \mathbb{R}^n$ to local variables $\xb_s \in \mathbb{R}^{n_s}$ associated with the corresponding subsystem. 
Note that the centralized model \eqref{LP_model} is equivalent to \eqref{LP_model_decomposed}, which can be decomposed into many separable component subproblems by relaxing the consensus constraint \eqref{DO_model_consensus}.

Existing distributed algorithms \cite{inaolaji2023distributed, muhlpfordt2021distributed, mhanna2018adaptive, dall2013distributed, pinto2020distributed} for solving \eqref{LP_model_decomposed} frequently rely on optimization solvers to solve component subproblems. For instance, the branch subproblem in ADMM \cite{mhanna2018adaptive} requires optimization solvers, whereas the generator and bus subproblems have closed-form solutions.
Calling optimization solvers at every iteration of distributed algorithms could be time-consuming and hinder their scalability, as highlighted in \cite{pinto2020distributed}, which employs ADMM to solve the component-wise distributed multi-phase OPF model. 
We tackle this issue by proposing a solver-free ADMM in the following section, which will be compared against the ADMM used for solving \eqref{LP_model_decomposed} in Section \ref{sec:experiments}.

\section{A GPU-accelerated distributed algorithm} \label{sec:proposed}
In this section we present a GPU-accelerated distributed algorithm which is built upon a solver-free ADMM. 
In contrast to the existing distributed algorithms discussed in Section \ref{sec:dist_opf}, the proposed distributed algorithm is solver-free, as each component subproblem has a closed-form solution expression through matrix operations. Eliminating the need for an optimization solver at every iteration can result in a significant reduction in computational time per iteration.
Therefore, our approach can offer enhanced scalability, particularly when GPUs are effectively leveraged for conducting efficient large-scale matrix operations through parallel computation.
\subsection{Reformulations}
First, with the component-wise decomposition strategy, we rewrite \eqref{LP_model} as
\begin{subequations}
\label{OurModel}
\begin{align}
\min \ & \cb^{\top} \xb \\
\mbox{s.t.} \ & \Ab_s \xb_s = \bb_s, \ \forall s \in [S], \label{OurModel_subconst}  \\
& \Bb_s \xb = \xb_s, \ \forall s \in [S], \label{OurModel_consensus} \\
& \xbl \leq \xb \leq \xbu. \label{OurModel_bound} 
\end{align}
\end{subequations}
The reformulation \eqref{OurModel} is key to run our algorithm on GPUs by isolating the bound constraints \eqref{OurModel_bound} from the local equality constraints \eqref{OurModel_subconst}. 
It is worth noting that \eqref{OurModel} is different from but equivalent to the model \eqref{LP_model_decomposed}. 

Practically speaking, the role of a system operator is reduced from computing \eqref{LP_model} in a centralized manner to managing the entire network within a given operational bounds in \eqref{OurModel}.
Concurrently, multiple agents are entrusted with the control and management of their respective components, drawing upon their own distinct sets of data.
In this distributed setting, the operator and the agents communicate iteratively until a global optimum is achieved. 
To this end, for every $t$-th iteration, the operator receives local solutions $\lambdab_s^{(t)}$ and $\xb_s^{(t)}$ from every agents $s \in [S]$ and solves the following optimization problem:
\begin{align}
\min_{\xb \in [\xbl, \xbu]} \cb^{\top}\xb + \sum_{s=1}^S \Big\{ \langle \lambdab_s^{(t)}, \Bb_s \xb  \rangle + \frac{\rho}{2} \|\Bb_s \xb - \xb_s^{(t)} \|^2 \Big\}, \label{ADMM-1} 
\end{align}
which computes an optimal solution $\xb^{(t+1)}$.
Then, each agent controlling a component $s \in [S]$ receives $\xb^{(t+1)}$ from the operator and solves the following optimization problem:
\begin{align}
\min_{\xb_s \in \mathbb{R}^{n_s}} \ & - \langle \lambdab_s^{(t)}, \xb_s  \rangle + \frac{\rho}{2} \|\Bb_s \xb^{(t+1)} - \xb_s \|^2  \nonumber \\     
\mbox{s.t.} \ & \Ab_s \xb_s = \bb_s, \label{ADMM-2} 
\end{align}
and update the dual variables
\begin{align}
\lambdab^{(t+1)}_s = \lambdab^{(t)}_s + \rho (\Bb_s \xb^{(t+1)} - \xb_s^{(t+1)}).   \label{ADMM-3} 
\end{align}  
Note that \eqref{ADMM-1}--\eqref{ADMM-3} collectively constitute one iteration of the conventional ADMM algorithm for solving \eqref{OurModel}, derived from the following augmented Lagrangian formulation:
\begin{align*}
\max_{ \{\lambdab_s \in \mathbb{R}^{m_s} \}_{s} }  \min \ & \cb^{\top} \xb + \sum_{s=1}^S  \langle \lambdab_s, \Bb_s \xb - \xb_s \rangle + \frac{\rho}{2} \|\Bb_s \xb - \xb_s \|^2   \\
\mbox{s.t.} \ & \eqref{OurModel_subconst}, \eqref{OurModel_bound}.
\end{align*}
where $\lambdab_s \in \mathbb{R}^{m_s}$ is a dual variables vector associated with the consensus constraint \eqref{OurModel_consensus}, and $\rho>0$ is the penalty parameter.

\subsection{Closed-form solution expressions}
In this section we present closed-form solution expressions for subproblems \eqref{ADMM-1} and \eqref{ADMM-2}.

First, we observe that the subproblem \eqref{ADMM-1} is separable over each element of $\xb$. 
To see this, we introduce a set $\Ic_{si} := \{ j \in [n_s] : (\Bb_s)_{j,i} = 1 \}$ for all $s \in [S]$ and $i \in [n]$.
With this set, \eqref{ADMM-1} can be decomposed into $n$-subproblems, each $i \in [n]$ of which is given by
\begin{align}
\min_{x_i \in [\underline{x}_i, \overline{x}_i]} \ \big(\cb + \sum_{s=1}^S \Bb^{\top}_s \lambdab_s^{(t)} \big)_i  x_i + \frac{\rho}{2} \sum_{s=1}^S  \sum_{j \in \Ic_{si}} \big(x_i - (\xb^{(t)}_s)_j \big)^2, \nonumber
\end{align}
which is an 1-D optimization problem with a convex quadratic objective function and a bound constraint.
The closed-form solution of the problem is given by
\begin{align}
x_i^{(t+1)} = \min \ \{ \max \{ \widehat{x}_i, \underline{x}_i \}, \ \overline{x}_i \} \label{closed_1}
\end{align}
where
\begin{align}
\widehat{x}_i = \frac{-1}{\rho \sum_{s=1}^S | \Ic_{si}| } \Big\{  \big(\cb + \sum_{s=1}^S \Bb^{\top}_s \lambdab_s^{(t)} \big)_i - \sum_{s=1}^S \sum_{j \in \Ic_{si}} (\xb^{(t)}_s)_j   \Big\}.  \nonumber
\end{align}
Second, for each subsystem $s \in [S]$, we rewrite the subproblem \eqref{ADMM-2} as the following quadratic program:
\begin{subequations}
\label{quadratic_problem}
\begin{align}
\min \ & \frac{1}{2} \xb^{\top}_s \Qb_s \xb_s + \langle \db^{(t)}_s, \xb_s \rangle  \\
\mbox{s.t.} \ & \ \Ab_s \xb_s = \bb_s, 
\end{align}
\end{subequations}
where $\Qb_s=\rho \Ib_{n_s}$ and $\db^{(t)}_s=-\rho \Bb_s \xb^{(t+1)} -\lambdab_s^{(t)}$.
Without loss of generality, we assume that $\Ab_s$ is a full row-rank matrix.
If $\Ab_s$ is not given as a full row-rank matrix, row reduction techniques can be applied to $\Ab_s \xb_s = \bb_s$ as a preprocessing step to rectify this. 
Since $\Ab_s$ is a full row-rank matrix and $\Qb_s$ is a symmetric positive definite matrix, a closed-form solution expression can be readily derived.
To see this, we introduce dual variables $\mub_s$ associated with $\Ab_s \xb_s = \bb_s$. Then the Lagrangian function is given by
\begin{align}
\max_{\mub \in \mathbb{R}^{m_s} } \min_{\xb_s \in \mathbb{R}^{n_s}} \frac{1}{2} \xb^{\top}_s \Qb_s \xb_s + \langle \db^{(t)}_s, \xb_s \rangle  + \langle \mub_s , \Ab_s \xb_s - \bb_s \rangle.\nonumber
\end{align}
For given $\mub_s \in \mathbb{R}^{m_s}$, one can derive from the first optimality condition that
\begin{align}
\xb_s = \frac{1}{\rho}   (-\db^{(t)}_s - \Ab^{\top}_s\mub_s). \nonumber
\end{align}
Plugging $\xb_s$ into $\Ab_s \xb_s = \bb_s$ yields
\begin{align}
\mub = - \rho (\Ab_s \Ab^{\top}_s)^{-1} \bb_s -  (\Ab_s \Ab^{\top}_s)^{-1} \Ab_s \db^{(t)}_s.\nonumber
\end{align}
Therefore, we have
\begin{subequations}
\label{closed_2}
\begin{align}
\xb^{(t+1)}_s = \frac{1}{\rho} \overline{\Ab}_s \db^{(t)}_s + \overline{\bb}_s \label{closed_2_1}
\end{align}
where
\begin{align}
& \overline{\Ab}_s := \Ab^{\top}_s (\Ab_s \Ab^{\top}_s)^{-1} \Ab_s - \Ib_{n_s} \\
& \overline{\bb}_s := \Ab^{\top}_s (\Ab_s \Ab^{\top}_s)^{-1} \bb_s
\end{align}
\end{subequations}
Note that both $\overline{\Ab}_s$ and $\overline{\bb}_s$ are computed only once at a preprocessing step.


\subsection{Termination criterion}
We use the standard primal and dual residuals to determine the termination of Algorithm \ref{alg:admm}, which are given as
\begin{align}
 \text{pres}^{(t)} \leq  \epsilon_{\text{prim}}^{(t)}, \ \   \text{dres}^{(t)} \leq \epsilon_{\text{dual}}^{(t)}, \label{termination}
\end{align}
where
\begin{align*}
& \textstyle \text{pres}^{(t)} := \sqrt{\sum_{s=1}^S \| \mathbf{B}_s \xb^{(t)} - \xb_s^{(t)}\|^2}  , \\
& \textstyle  \text{dres}^{(t)} :=  \rho \sqrt{ \sum_{s=1}^S \| \mathbf{B}^{\top}_s(\xb^{(t)}_s - \xb^{(t-1)}_s) \|^2 } , \\
& \textstyle \epsilon_{\text{prim}}^{(t)}:= \epsilon_{\text{rel}} \max \Big\{  \sqrt{\sum_{s=1}^S \|\Bb_s \xb^{(t)} }\|^2, \ \sqrt{\sum_{s=1}^S \| \xb^{(t)}_s \|^2 } \Big \}, \\
& \textstyle \epsilon_{\text{dual}}^{(t)}:= \epsilon_{\text{rel}}     \sqrt{\sum_{s=1}^S \|\Bb_s^{\top} \lambdab^{(t)}_s }\|^2 , \\
\end{align*} 
and $\epsilon_{\text{rel}}$ is a tolerance level (e.g., $10^{-2}$ or $10^{-3}$).

\subsection{The proposed distributed algorithm}
The proposed algorithm is summarized in Algorithm \ref{alg:admm}.
After initializing dual and local solutions as in line 1, we precompute the matrices and vectors used for local updates at each subsystem $s \in [S]$ as in lines 2-3. 
Then, the proposed iterative algorithm runs as in lines 4-10 until a termination criterion \eqref{termination} is satisfied.
For every iteration $t$ of the algorithm, the computational steps are taken to update a global solution as in line 5, local solutions as in line 7, and dual solutions as in line 8.

\begin{algorithm}[h!]
\caption{A solver-free ADMM for solving a component-wise distributed multi-phase OPF model \eqref{OurModel}. }\label{alg:admm}
\begin{algorithmic}[1]
\State Initialization: $t \gets 1$, $\lambdab^{(t)}_s$, and $\xb^{(t)}_s$  for all $s \in [S]$
\Statex Precomputation:
\For { $s \in [S]$ }{ in parallel}
\begin{align*}
  & \overline{\Ab}_s := \Ab^{\top}_s (\Ab_s \Ab^{\top}_s )^{-1}  \Ab_s - \Ib_{n_s} \\
  & \overline{\bb}_s := \Ab^{\top}_s (\Ab_s \Ab^{\top}_s )^{-1}\bb_s
\end{align*}
\EndFor
\While{\eqref{termination} is not satisfied}
\State Compute $\xb^{(t+1)}$ by \eqref{closed_1} \Comment{Global Update}
\For { $s \in [S]$ }{ in parallel}
\State Compute $\xb^{(t+1)}_s$ by \eqref{closed_2} \Comment{Local Update}
\State Compute $\lambdab^{(t+1)}_s$ by \eqref{ADMM-3} \Comment{Dual Update}
\EndFor 
\EndWhile 
\end{algorithmic} 
\end{algorithm}

The proposed algorithm is an ADMM algorithm composed of subproblems that admit closed-form solution expressions.
The sequence of iterates generated by ADMM converges to an optimal solution with sublinear rate \cite{he20121}.
A few acceleration schemes exist that could reduce the total number of iterations, including residual balancing \cite{wohlberg2017admm} and multiple local updates \cite{ryu2022differentially}.
In this paper, instead of proposing new acceleration schemes to reduce the total number of iterations, we focus on improving computation time per iteration by running the algorithm on GPUs. 
In the next section we describe the implementation of Algorithm \ref{alg:admm} for efficient execution on GPU

\section{Implementation Details for the GPU computations} \label{sec:implementation}
To enable GPU acceleration, we implement the proposed distributed algorithm in Julia with the CUDA.jl package \cite{besard2018effective} for the NVIDIA GPU programming.
\subsection{Motivation}
A GPU has been used for conducting large matrix operations mainly because of its ability of conducting such operations in parallel using multiple threads.
For example, when conducing $\Ab \xb$ where $\Ab \in \mathbb{R}^{m \times n}$ and $\xb \in \mathbb{R}^{n}$, each thread of a GPU conducts a  $\mathbf{a}^{\top}_i \xb$ where $\mathbf{a}^{\top}_i$ is $i$-th row of $\Ab$.
Since Algorithm \ref{alg:admm} is composed of a set of matrix operations as in \eqref{closed_1}, \eqref{closed_2}, and \eqref{ADMM-3}, we aim to effectively leverage GPU for further acceleration in per-iteration computing time.

\subsection{Row reduction}
Algorithm \ref{alg:admm} assumes that $\Ab_s$ is a full row rank matrix, which is used for constructing local feasible region $\mathcal{X}_s:=\{\xb_s : \Ab_s \xb_s = \bb_s \}$ for every component $s \in [S]$. 
However, in practice, $\Ab_s$ might not have full row rank. In such case, we apply a row reduction technique to put the augmented matrix $[\Ab_s | \bb_s]$ in a reduced row echelon form, leading to a full row rank matrices.
We note that the size of $\Ab_s$ is small (e.g., see Table \ref{tab:size_distributed}) as we consider the component-wise decomposition strategy, therefore, the row reduction process is very fast, which also can be done in parallel across multiple components. 

\subsection{Global and dual updates in Algorithm \ref{alg:admm}} \label{sec:global_dual}
The global update \eqref{closed_1} and dual update \eqref{ADMM-3} can be efficiently conducted by leveraging sparse matrix operations.
To this end, we first rewrite the consensus constraints \eqref{OurModel_consensus} as 
\begin{align}
\begin{bmatrix}
\Bb_1 \\    
\vdots \\
\Bb_S \\ 
\end{bmatrix}
\xb =
\begin{bmatrix}
\xb_1 \\    
\vdots \\
\xb_S \\ 
\end{bmatrix}
\ \Leftrightarrow \ \Bb \xb = \zb,
\end{align}
where $\Bb$ and $\zb$ are concatenated matrix and vector, respectively.
Then the global update \eqref{closed_1} can be rewritten as 
\begin{align}
& \hat{\xb} \gets -\frac{1}{\rho} (\Bb^{\top} \Bb)^{-1} \cb + (\Bb^{\top} \Bb)^{-1} \Bb^{\top} (\zb^{(t)} - \frac{1}{\rho} \lambdab^{(t)} ), \nonumber \\
& \xb^{(t+1)} \gets \min \{ \max \{ \hat{\xb}, \xbl \}, \xbu \}, \label{global_sparse}
\end{align}
where $\zb=[\xb_1; \ldots ; \xb_S]$ and $\lambdab^t = [\lambdab_1; \ldots ; \lambdab_S]$. The dual update can be rewritten as 
\begin{align}
\lambdab^{(t+1)}   \gets \lambdab^{(t)} + \rho (\Bb \xb^{(t+1)} - \zb^{(t+1)}), \label{dual_sparse}
\end{align}
where $\zb^{(t+1)}$ is obtained from the local update \eqref{closed_2}.

Since the matrix $\Bb$ is sparse, we leverage sparse matrix operation to efficiently conduct both global and dual updates.
Specifically, we use \texttt{CuArray} supported by the CUDA.jl package for writing such matrices and vectors and precompute $(\Bb^{\top} \Bb)^{-1} \cb $ and $(\Bb^{\top} \Bb)^{-1} \Bb^{\top}$, before executing the ADMM iterations (i.e., before line 4 in Algorithm \ref{alg:admm}).
This process is very fast primarily because the matrix $\Bb^{\top} \Bb$ is diagonal, and its inverse is also diagonal, with each diagonal element being the reciprocal of the corresponding diagonal element of $\Bb^{\top} \Bb$.

\subsection{Local update in Algorithm \ref{alg:admm}}
With the component-wise decomposition, it is often the case that the size of the matrix $\Ab_s$ for each component $s \in [S]$ is very small.
In such situation, using \texttt{CuArray} as in global and dual updates described in Section \ref{sec:global_dual} may not improve the computation time for conducting matrix operations required in \eqref{closed_2} for the local update. To address this, we construct a kernel function that assigns the vector operations to each thread of a GPU and enable the computation in parallel.   

In CUDA, parallel programs (kernels) execute on a grid consisting of multiple blocks, each containing the same number of threads. Each thread has a unique identifier and serves as the smallest unit of parallel computation. All threads within a grid execute the same kernel simultaneously.
To leverage threads for conducting local update \eqref{closed_2} in parallel, we set the number of blocks to $S$ (i.e., the number of components).
For each block $s \in [S]$, we set the number of threads as a parameter (e.g., $T \in \{1, \ldots, 64\}$). 
Then each thread of the block $s$ is used for computing the $i$-th entry of $\xb^{(t+1)}_s$ in \eqref{closed_2}.

\subsection{Communication}
For communication between multiple GPUs within a computing cluster, we use the MPI.jl package \cite{byrne2021mpi}. For every iteration of Algorithm \ref{alg:admm}, it requires a communication of intermediate solutions, including (i) $\xb^{(t+1)}$ in line 5 sent from a server to all subsystems, and (ii) $\xb^{t+1}_s$ in line 7 an $\lambdab_s^{t+1}$ in line 8 sent from each subsystem to the server.
We note that the communication time between multiple GPUs is often higher than that between multiple CPUs because MPI requires to transfer data from GPU to CPU to communicate. 
In practice, however, multiple GPUs are often located at different computing resources, thus MPI cannot be used for communication.
In this case, a remote procedure call which enables communication between multiple platforms can be utilized.
With tRPC \cite{damania2023pytorch}, for example, the communication time between multiple GPUs is about the same as that between multiple CPUs, as reported in \cite{damania2023pytorch}.
This can rule out a potential concern of communication overhead due to the introduction of GPUs.


\section{Numerical Results} \label{sec:experiments}
In this section we aim to demonstrate how the proposed GPU implementation of Algorithm \ref{alg:admm} accelerates the computation compared with its CPU counterparts.

All numerical tests were performed on (i) Swing, a 6-node GPU cpomputing cluster (each node has 8 NVIDIA A100 40GB GPUs) and (ii) Bebop, a 1024-node CPU computing cluser (each node has 36 cores with Intel Xeon E5-2695v4 processors and 128 GB DDR4 of memory) at Argonne National Laboratory.

\subsection{Experimental settings}
For our demonstration, we use IEEE test instances \cite{schneider2017analytic} with 13, 123, and 8500 buses.
In these instances, the three-phase loads at each bus are given as either wye or delta connected and each load is labeled as either constant power, current, or impedance load. 
These features are taken into a consideration by our mathematical model in Section \ref{sec:math_form}.
The size of matrix $\Ab$ in \eqref{LP_model} for each instance is reported in Table \ref{tab:size_A}.
Note that the number of columns in this table is the size of global variables $\xb$ in \eqref{OurModel}.

\begin{table}[!h]
\centering
\caption{The number of rows and columns of $\Ab$ in \eqref{LP_model} for the test instances.}
\label{tab:size_A}
\begin{tabular}{ccc}
\hline  
IEEE13 &  IEEE123& IEEE8500 \\ \hline
(456, 454)  & (1834, 1834) & (86114, 87285)  \\
\hline
\end{tabular}
\end{table} 

For the distributed setting, we construct $S$ subsystems in \eqref{OurModel} based on the network decomposition by its network components such as buses, transformers, and branches, as in \cite{mhanna2018adaptive}.
Specifically, we construct a graph with a set of lines, each of which represents either branch or transformer line, and a set of nodes, each of which represents either bus or a node connecting to a transformer line. 
Then we construct a set of subproblems related to each component of the graph.
Based on our observation that it is often the case that the subproblems related to leaf nodes and their connected lines are much smaller than the other subproblems, we construct a set of subsystems, each of which is constructed by combining a leaf node to its connecting line and another set of subsystems, each of which is constructed by the remaining individual node or line.  
We report the total number $S$ of components in Table \ref{tab:subsystems}.
For parallel computation utilizing $N$ nodes or cores, such as CPUs, GPUs, or threads of a GPU, we distribute $S$ subsystems nearly evenly, assinging each one to a distinct node $n \in [N]$.

\begin{table}[!h]
  \centering
  \caption{Total number $S$ of components.}
  \label{tab:subsystems}
  \begin{tabular}{c|ccc}
  \hline
         & IEEE13 &   IEEE123 & IEEE8500 \\ \hline    
  \# of nodes   & 29&  147 & 11932 \\         
  \# of lines   & 28&    146 & 14291 \\         
  \# of leaf nodes & 7 &   43 & 1222 \\ 
  $S$           & 50&    250 & 25001 \\ \hline
  \end{tabular}  
\end{table}

The size of each compoenent subproblem $s \in [S]$ can be measured by $m_s$ and $n_s$ which are the number of rows and columns of $\Ab_s$ in \eqref{OurModel}. 
Since the number of constraints and variables of each subproblem depends on the number of phases, the instance like IEEE 8500 \cite{5484381} that has smaller number of phases at each component of the network has smaller size of subproblems as reported in Table \ref{tab:size_distributed}.

\begin{table}[!h]
\centering
\caption{The size of component subproblem measured by $m_s$ and $n_s$ for $s \in [S]$}
\label{tab:size_distributed}
\begin{tabular}{l|ccc}
\hline
                        & IEEE13 &   IEEE123 & IEEE8500 \\ \hline 
Min $(m_1,\ldots,m_S)$ & $4$  &  $2$& $2$\\ 
Max $(m_1,\ldots,m_S)$ & $22$ &   $42$& $18$ \\
Mean $(m_1,\ldots,m_S)$ & $9.08$& $7.34$ & $3.44$ \\ 
Stdev $(m_1,\ldots,m_S)$ & $4.42$&  $4.43$ & $2.66$\\ 
Sum $(m_1,\ldots,m_S)$ & $453$&   $1834$ & $86108$\\  \hline
Min $(n_1,\ldots,n_S)$ & $8$  &   $4$& $4$\\ 
Max $(n_1,\ldots,n_S)$ & $34$ &   $57$&$24$ \\
Mean $(n_1,\ldots,n_S)$ & $16.1$&  $13.16$ & $6.69$\\ 
Stdev $(n_1,\ldots,n_S)$ & $5.14$&  $6.5$ & $3.21$\\ 
Sum $(n_1,\ldots,n_S)$ & $805$ &   $3289$ & $167394$\\     
\hline
\end{tabular}  
\end{table}

As a default setting for Algorithm \ref{alg:admm}, we set $\epsilon_{\text{rel}}=10^{-3}$ in \eqref{termination}, and $\rho=100$.
As an initial point to start (in line 1 of Algorithm \ref{alg:admm}), for each $s \in [S]$, we set $\lambdab^1_s \gets 0$ and set each element of $\xb^1_s$ as either (i) zero if it does not have bounds, (ii) the average value of lower and upper bounds if it has bounds, and (iii) one if it is related to the voltage magnitude.

\subsection{Performance comparison using multiple CPUs in parallel}
In this section, we compare the performance of Algorithm \ref{alg:admm} against a benchmark approach that also uses ADMM to solve model \eqref{LP_model_decomposed}, using the same number of CPU cores. Notably, GPU acceleration is not utilized for Algorithm \ref{alg:admm} in this comparison.

Specifically, one iteration of the benchmark ADMM consists of: (i) a global update, $x_i^{t+1} \gets \widehat{x}_i$, with $\widehat{x}_i$ obtained from \eqref{ADMM-1}; (ii) a local update, $\xb^{(t+1)}_s$ is computed by solving the quadratic model \eqref{quadratic_problem} subject to the bound constraints $\xb_s \in [\xbl_s, \xbu_s]$, defined for every component $s \in [S]$; and (iii) a dual update according to \eqref{ADMM-3}.
It is worth noting that the computation times spent on the global and dual updates for both approaches will be similar because those updates are composed of a set of matrix operations. 
However, the time for conducting the local updates will be significantly different since the local update process of the benchmark approach requires optimization solvers for solving the subproblems (i.e., the quadratic model with bound constraints) defined for every component $s \in [S]$, while the local update process of our approach is just matrix operations. 
In what follow, we will compare both approaches based on (i) the average time required for the local update in each iteration of the ADMM methods, and (ii) the total time and number of iterations needed for convergence.

\begin{figure*}[!h]
\centering
\begin{subfigure}[b]{0.32\textwidth}
\centering  
\includegraphics[width=\textwidth]{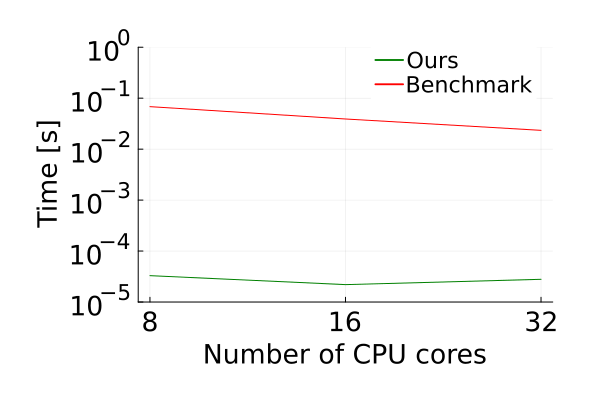}   
\includegraphics[width=\textwidth]{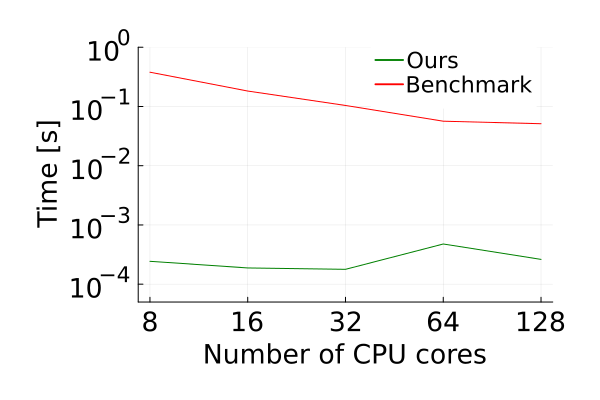}   
\includegraphics[width=\textwidth]{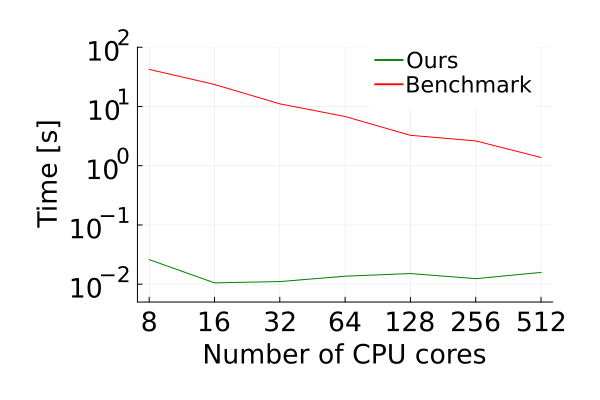}   
\caption{Local update}  
\label{fig:local_update_sum}
\end{subfigure}  
\begin{subfigure}[b]{0.32\textwidth}
\centering
\includegraphics[width=\textwidth]{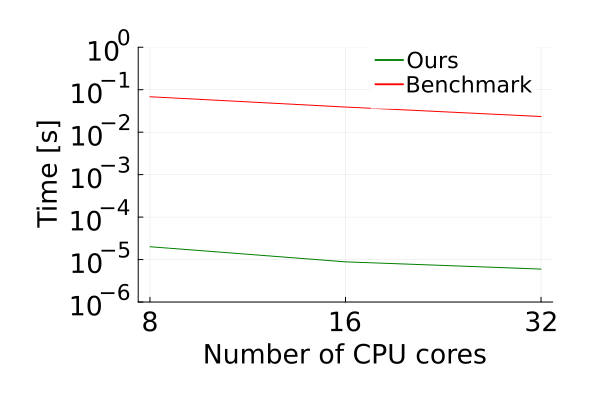}   
\includegraphics[width=\textwidth]{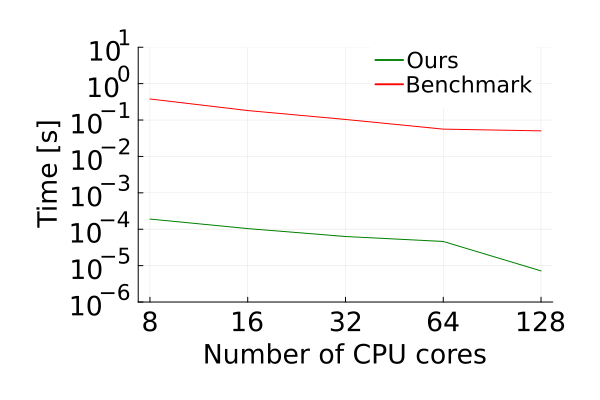}   
\includegraphics[width=\textwidth]{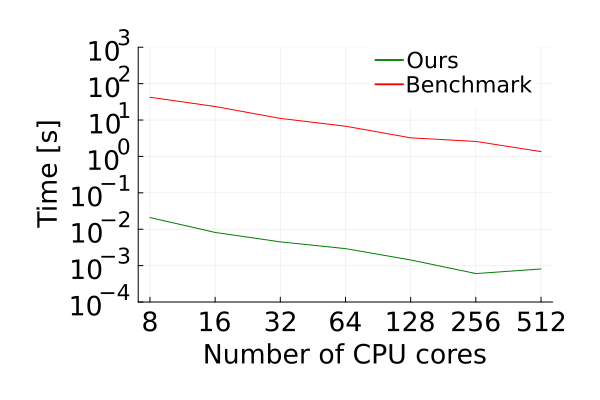}   
\caption{Computation}  
\label{fig:local_update_comp}
\end{subfigure}
\begin{subfigure}[b]{0.32\textwidth}
\centering  
\includegraphics[width=\textwidth]{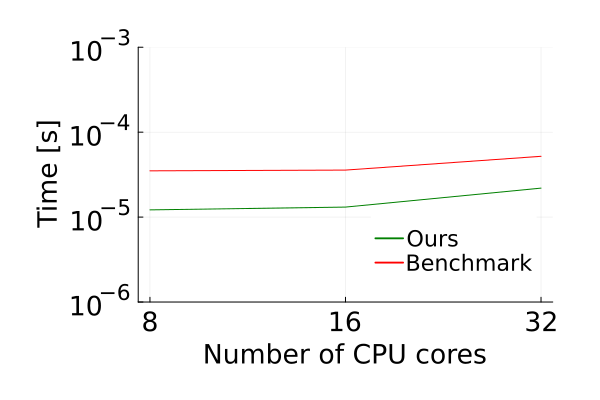}   
\includegraphics[width=\textwidth]{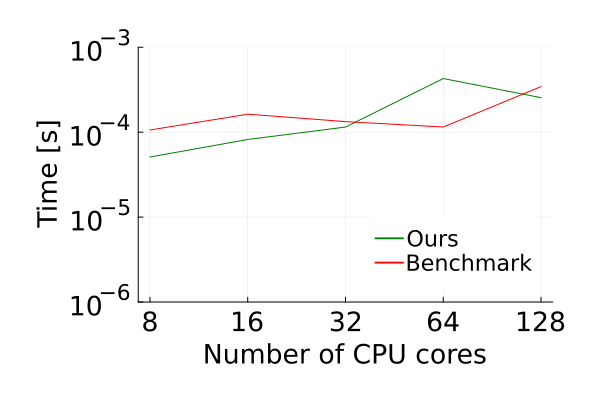}   
\includegraphics[width=\textwidth]{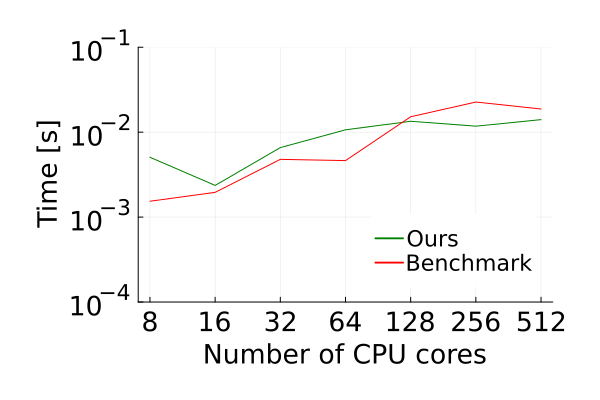}   
\caption{Communication}  
\label{fig:local_update_comm}
\end{subfigure}                  
\caption{
(a) Average wall-clock time consumed for the local update.
(b) Average computation time for solving the subproblems. 
(c) Average communication time. 
Note that the results from the first, second, and third rows are associated with the IEEE 13, 123, and 8500 instances, respectively. 
}
\label{fig:local_update}
\end{figure*}

\begin{table*}[h!]
\centering
\caption{Comparison of the two approaches in terms of total time and iterations consumed until convergence.}
\begin{tabular}{|c|r|r|r|r|r|r|r|r|r|}
\hline  
& \multicolumn{3}{c|}{IEEE 13} & \multicolumn{3}{c}{IEEE 123} & \multicolumn{3}{|c|}{IEEE 8500} \\ \hline
& \# CPUs  &  Time [s] & Iterations  & \# CPUs  &  Time [s] & Iterations  & \# CPUs & Time [s] & Iterations \\ \hline
Ours  & $16$ &  $4.91$    &  $944$ & $16$ &  $7.25$    &  $3496$  & $16$ & $668.30$ &  $15817$ \\
Benchmark  & $32$ &  $28.13$    &  $1064$ & $128$ & $169.67$    &  $3215$  & $512$ & $44720.11$ &  $26252$ \\ \hline
\end{tabular}
\label{table:total_time_iter}
\end{table*}
    
First, in Figure \ref{fig:local_update_sum}, we report the average wall-clock time in seconds consumed for conducting local update per iteration of Algorithm \ref{alg:admm} and the benchmark approach, respectively.
The local update process in this experiment is composed of (i) solving $S \in \{50, 250, 25001\}$ subproblems which are distributed over multiple CPUs for parallel computation, and (ii) communicating information between the aggregator and each subsystem.
The average computation and communication times consumed for a local update are presented in Figures \ref{fig:local_update_comp} and \ref{fig:local_update_comm}, respectively.
As the number of CPUs increases, the computation time consumed for solving the subproblems tends to decrease as shown in Figure \ref{fig:local_update_comp} while the communication time tends to increase as shown in Figure \ref{fig:local_update_comm}.
This is mainly because, as the number of CPU cores increases, each core will have less number of subproblems to solve, reducing the computation time, while more information needs to be communicated between each CPU core and a central aggregator, resulting in increasing communication time.
By adding the computation and communication times, we compute the total wall-clock time spent on the local update process, as illustrated in Figure \ref{fig:local_update_sum}. 
This demonstrates that the benchmark approach requires more CPUs to accelerate the local update process, whereas our algorithm is faster even with significantly fewer CPUs.

Next, in Table \ref{table:total_time_iter}, we present the total time and iterations needed by Algorithm \ref{alg:admm} and the benchmark, respectively, to meet the termination criterion, as well as the number of CPUs used for the experiment, determined by the results from Figure \ref{fig:local_update}. 
For the IEEE 13 and 123 instances, the number of iterations required by the two approaches are similar, but our approach is significantly faster than the benchmark in terms of the wall-clock time, approximately $7$ and $23$ times faster, respectively.
For the IEEE 8500 instance, our approach significantly improves both total time (approximately 67 times faster) and the number of iterations (1.66 times fewer) compared to the benchmark.
It is worth noting that our approach requires less CPU cores while the benchmark requires more CPU cores to accelerate the solving time. 
The performance of the benchmark could be enhanced by leveraging more CPU cores, but this requires leveraging multiple computing nodes in high performance computing cluster, which may be impractical.

\subsection{GPU acceleration}
In this section we numerically demonstrate that GPUs can significantly accelerate the computational performance of Algorithm \ref{alg:admm}.

\subsubsection{Convergence}
\begin{figure}[h!]
\centering    
\includegraphics[width=.45\linewidth]{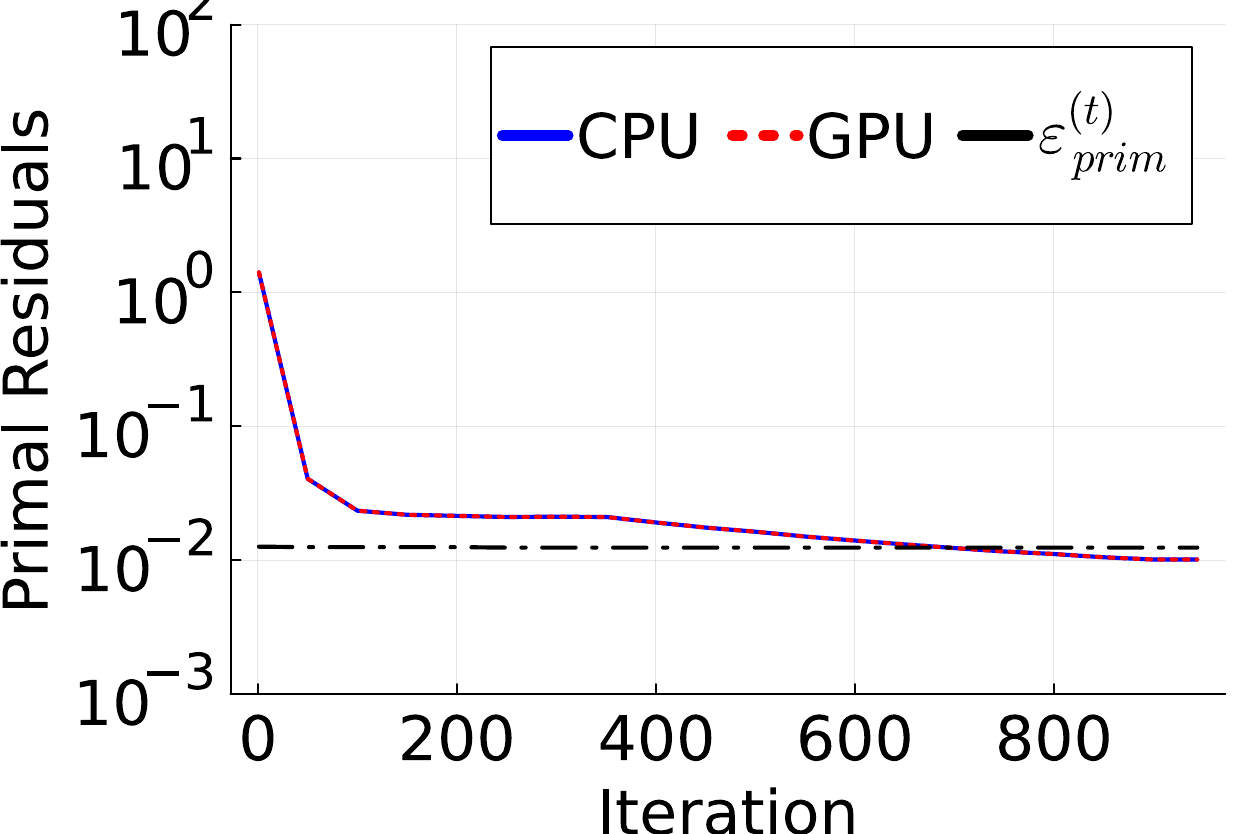}    
\includegraphics[width=.45\linewidth]{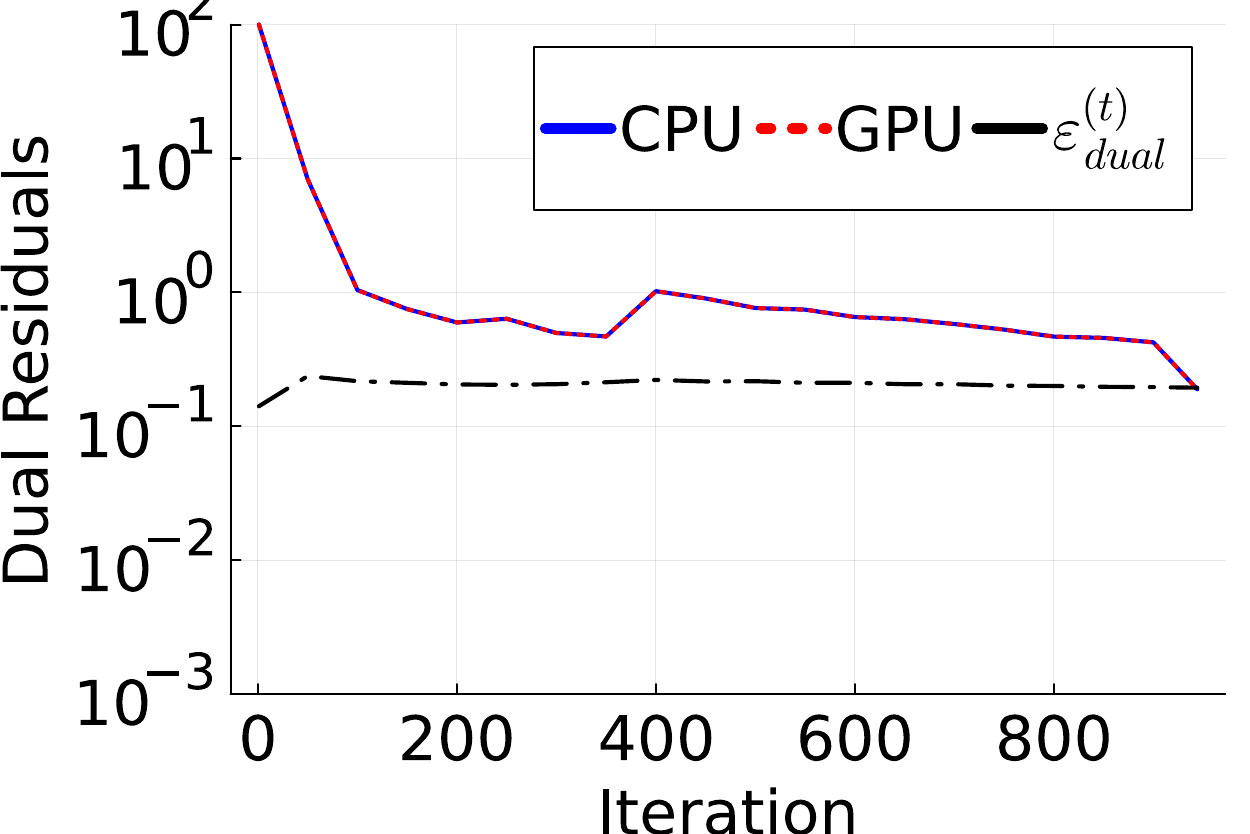}    
\caption{Primal (left) and dual (right) residuals at each iteration of Algorithm \ref{alg:admm} when a CPU and a GPU, respectively, is used for solving the IEEE 13 instance.}
\label{fig:residuals}
\end{figure}

We utilize GPU to accelerate the computation time for solving the component subproblems. Consequently, the total number of iterations required to reach convergence to an optimal solution should remain the same for both CPU and GPU computations. This is demonstrated in Figure \ref{fig:residuals}, which depicts the progression of the primal and dual residuals as the iterations of Algorithm \ref{alg:admm} increase for both CPU and GPU implementations using the IEEE 13-bus system. This confirms that the convergence behavior of Algorithm \ref{alg:admm} is consistent across both CPU and GPU computations.

\subsubsection{Computational performance}

\begin{figure*}[h!]
\centering
\begin{subfigure}[b]{0.3\textwidth}
\centering
\includegraphics[width=\textwidth]{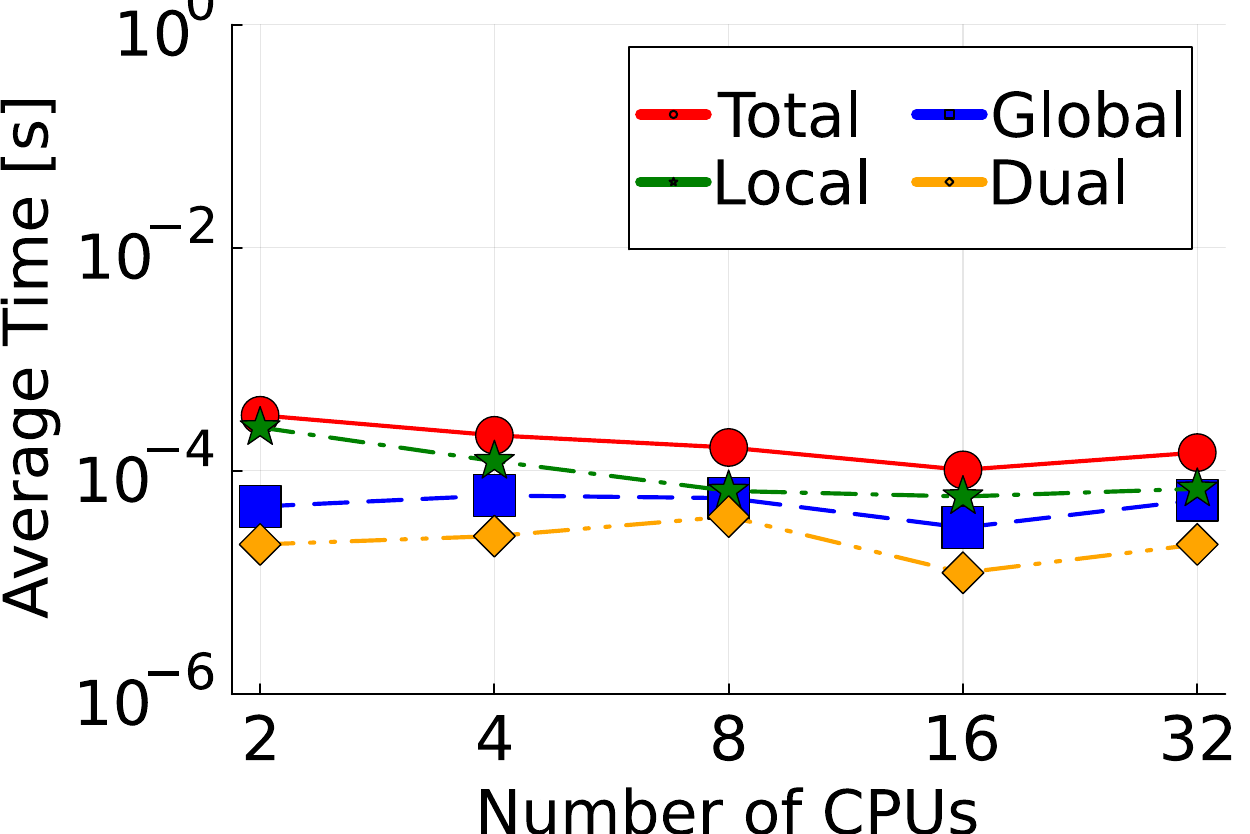} 
\includegraphics[width=\textwidth]{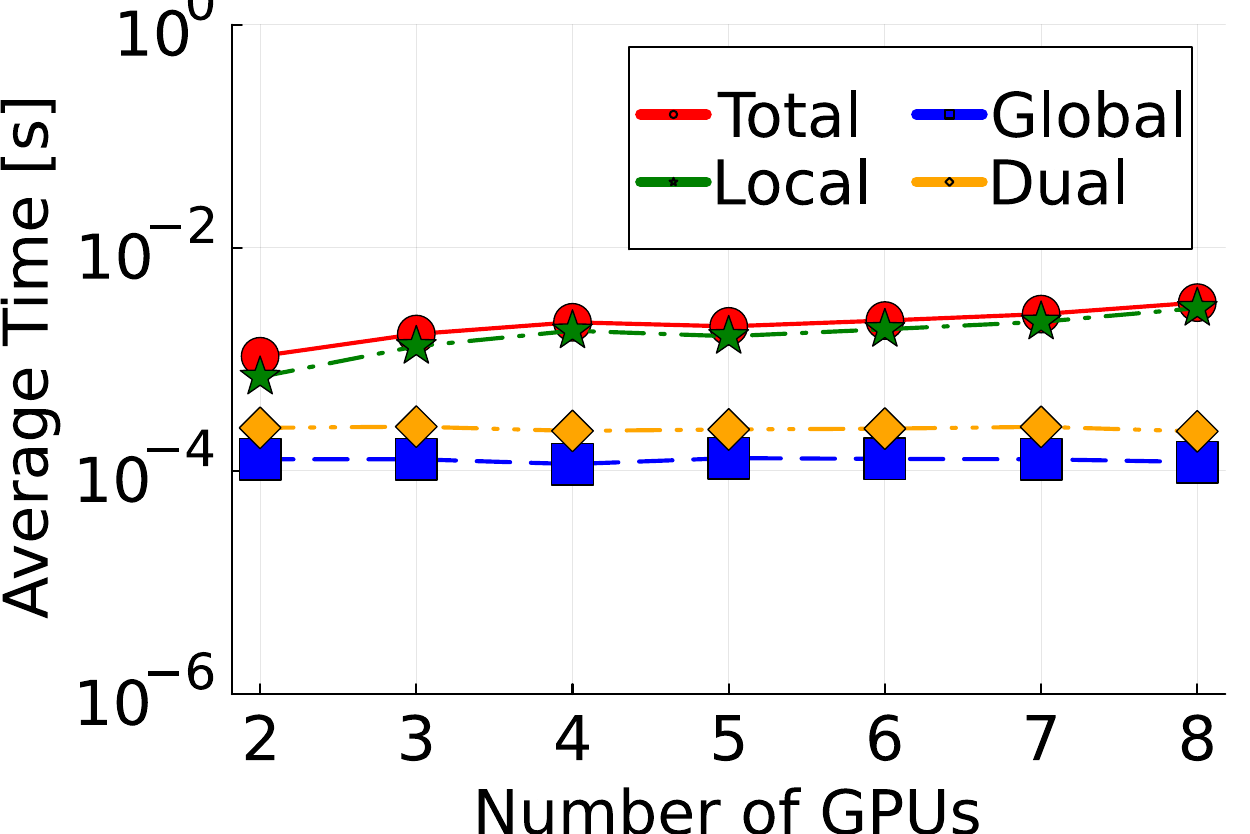}    
\includegraphics[width=\textwidth]{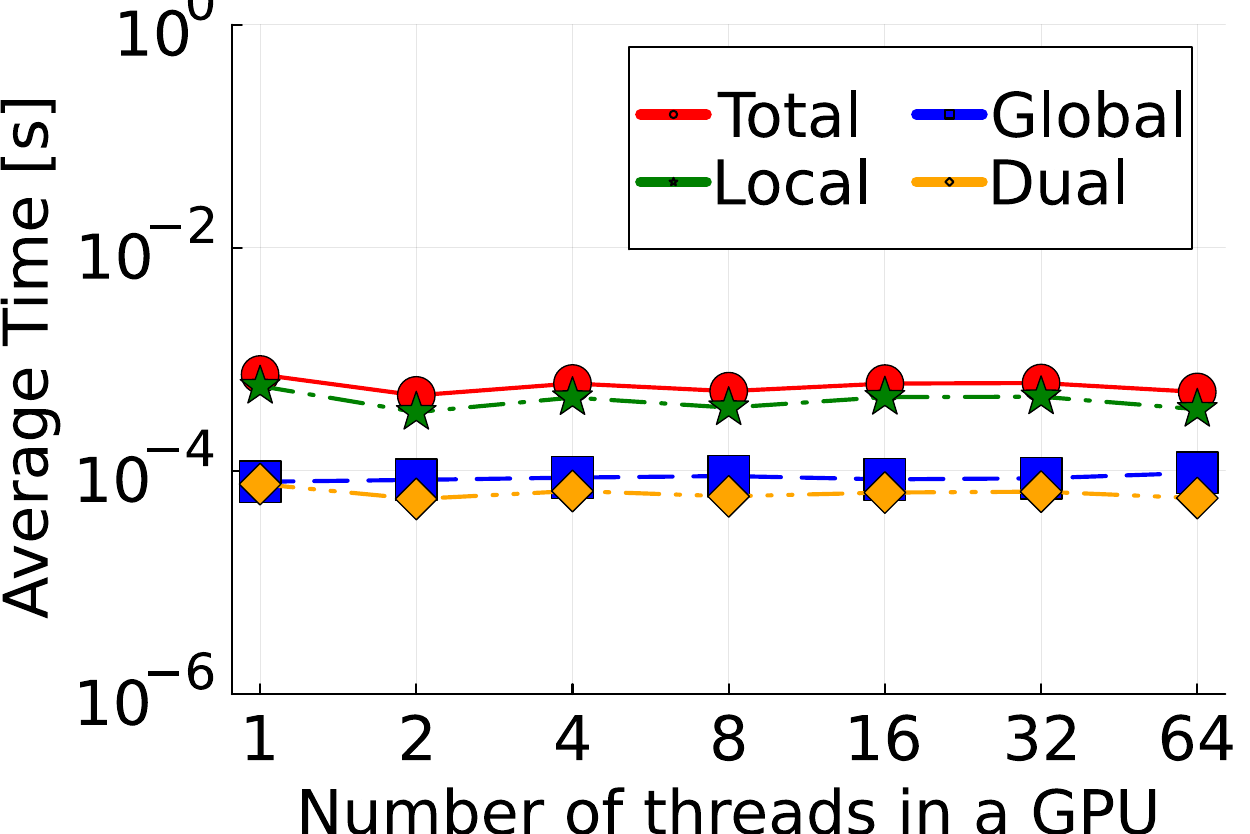}  
\caption{IEEE13}  
\end{subfigure} 
\begin{subfigure}[b]{0.3\textwidth}
\centering
\includegraphics[width=\textwidth]{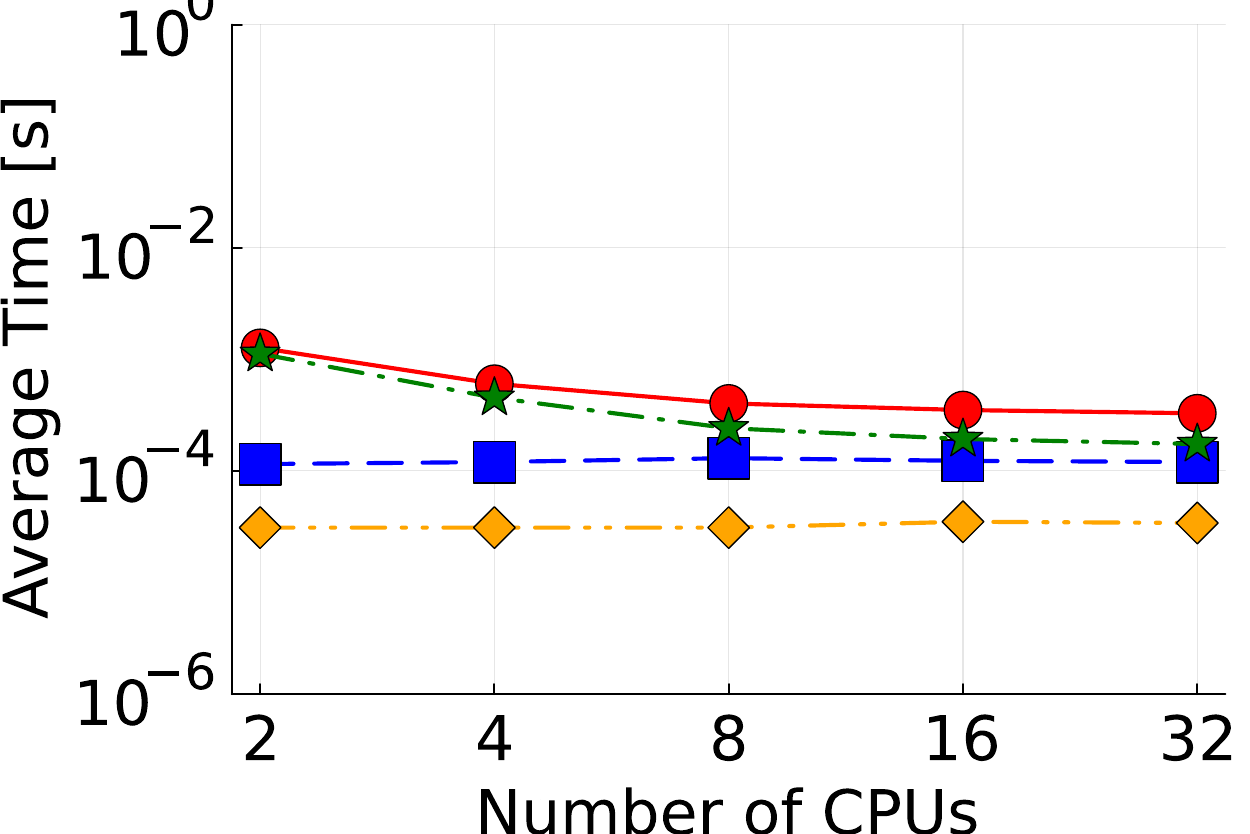} 
\includegraphics[width=\textwidth]{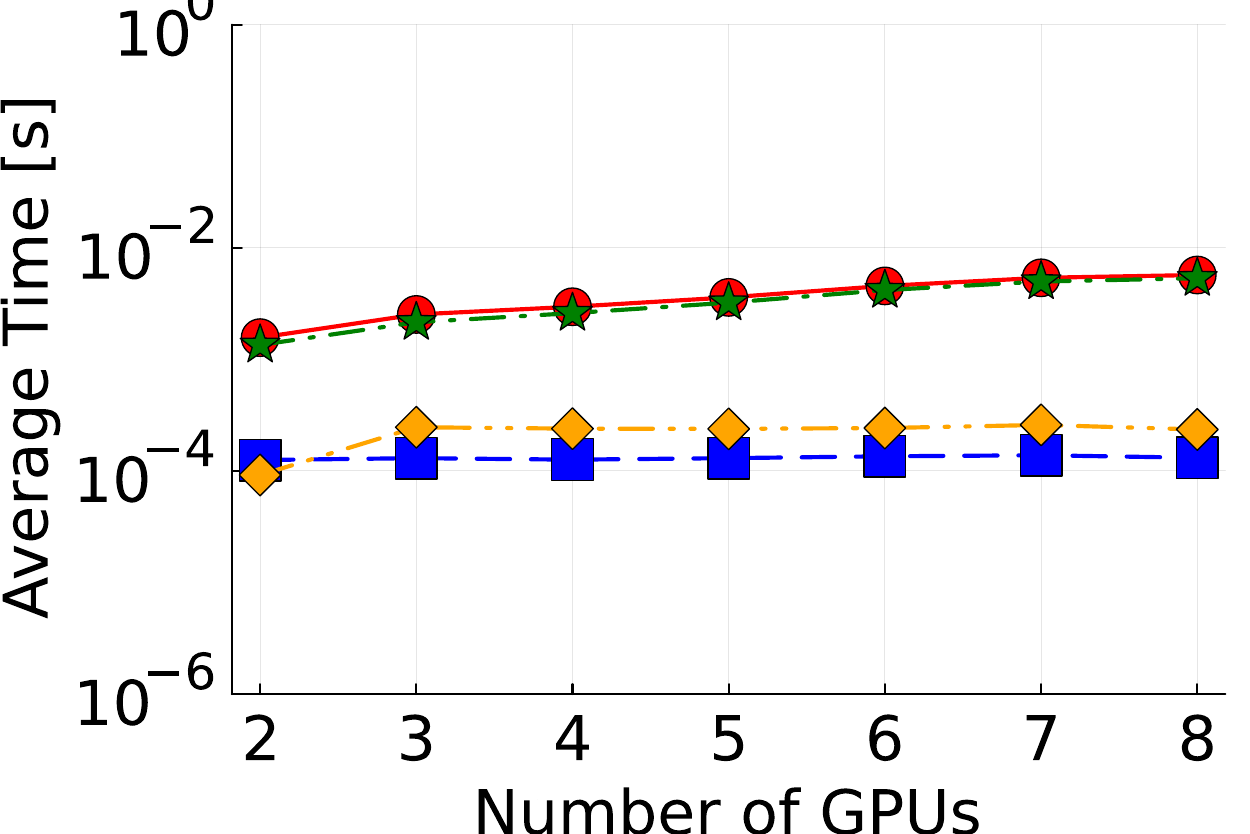}  
\includegraphics[width=\textwidth]{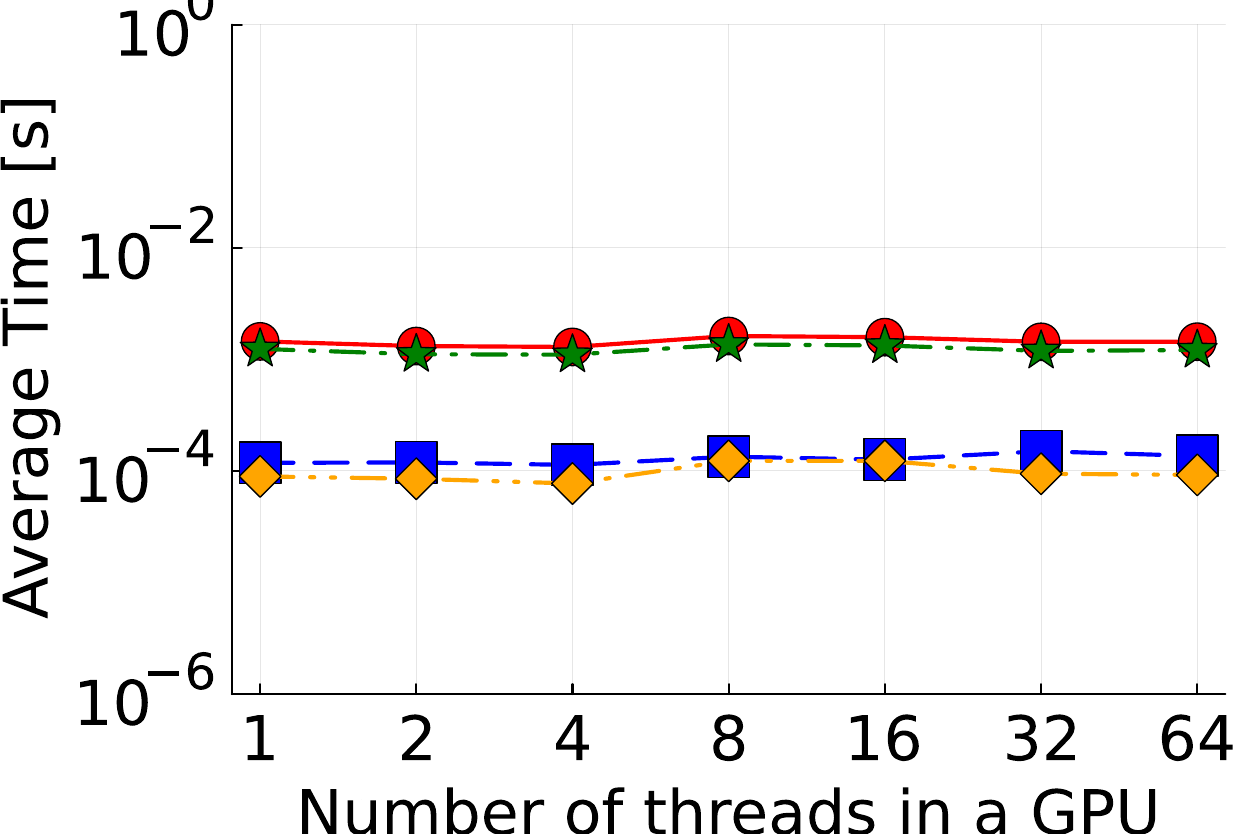}  
\caption{IEEE123}  
\end{subfigure}
\begin{subfigure}[b]{0.3\textwidth}
\centering  
\includegraphics[width=\textwidth]{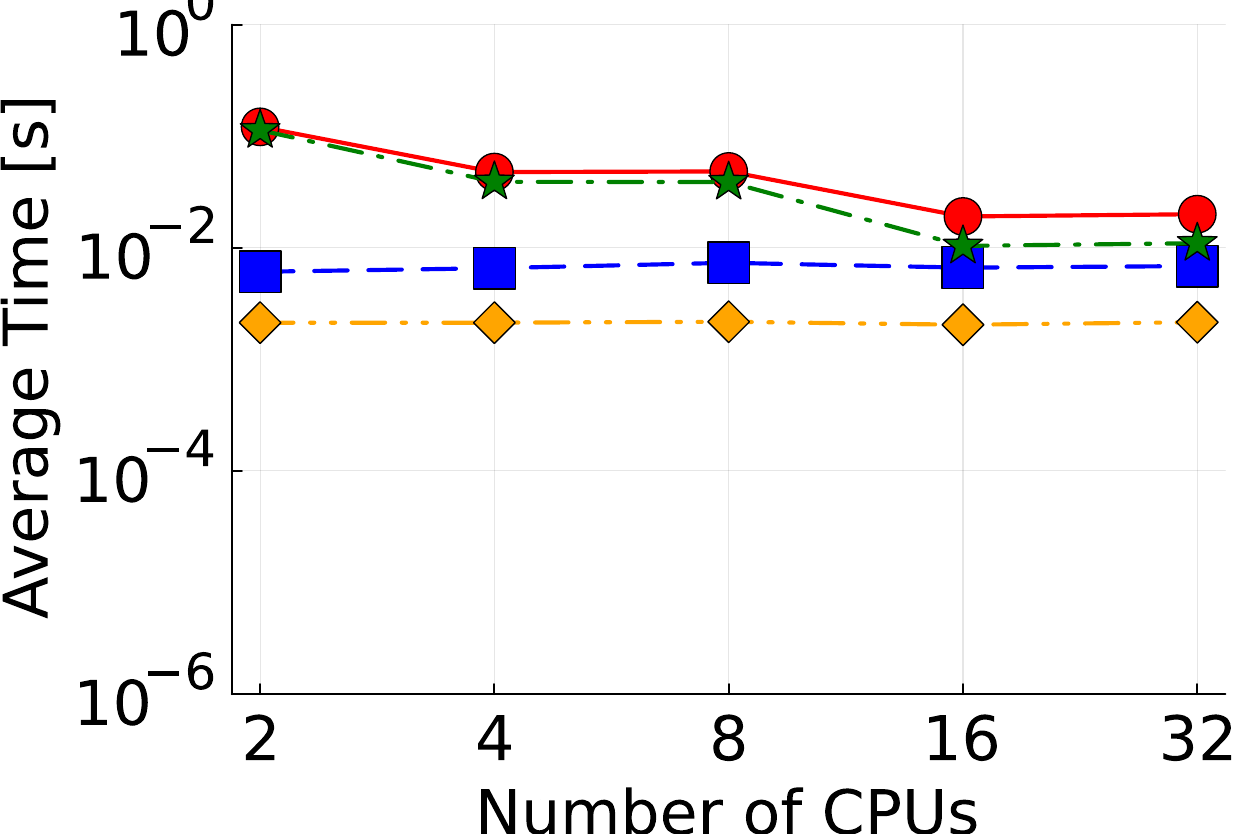}
\includegraphics[width=\textwidth]{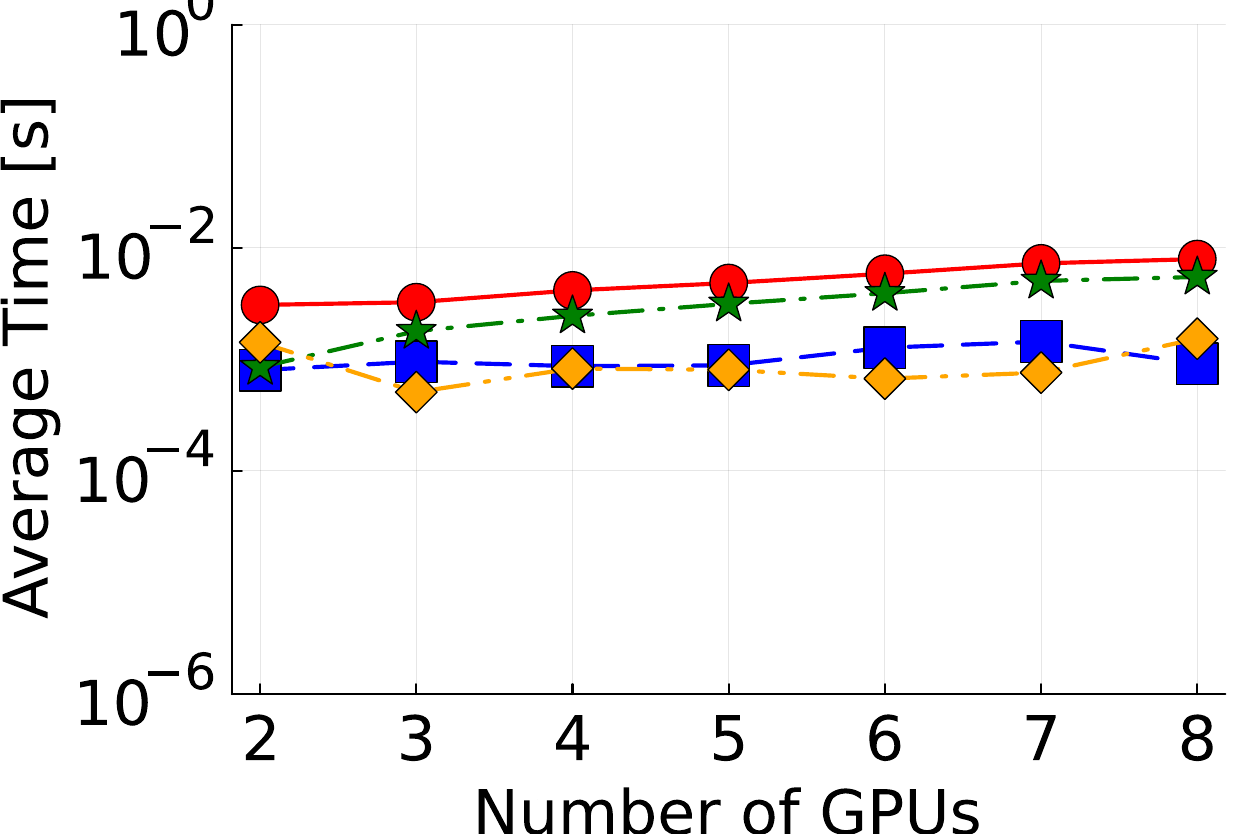}  
\includegraphics[width=\textwidth]{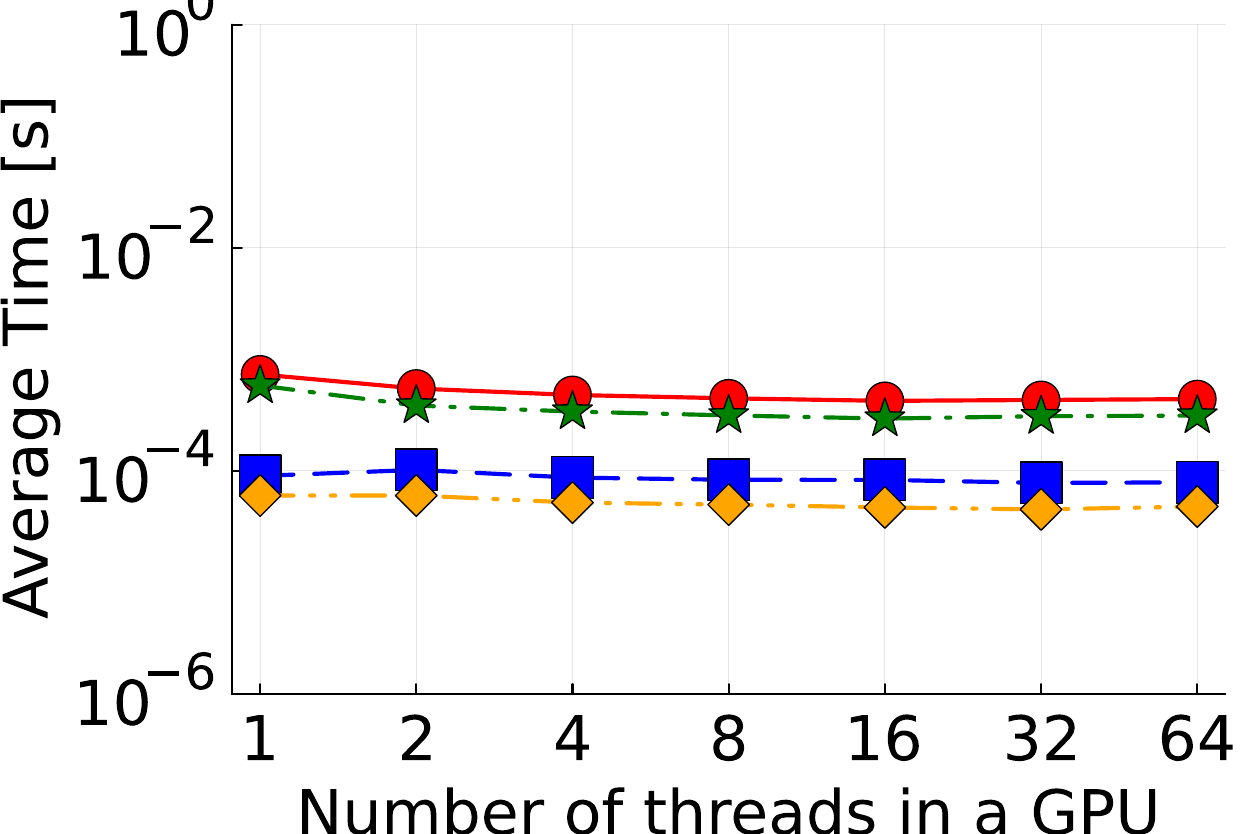} 
\caption{IEEE8500}  
\end{subfigure}          
\caption{Average time for conducting global, local, and dual updates per iteration and their summation (referred as total) when multiple CPUs (top), GPUs (middle), and threads in a GPU (bottom), respectively. Both multiple CPUs and GPUs computations are done in parallel.}
\label{Figure:cpus_gpus}
\end{figure*}

Given the equivalence in the number of iterations required for Algorithm \ref{alg:admm} to converge to an optimal solution when executed on both CPU and GPU platforms, it becomes imperative to enhance the efficiency of computation per iteration. In pursuit of this objective, we conduct parallel computation employing multiple CPUs and GPUs, respectively.
 
In the first row of Figure \ref{Figure:cpus_gpus}, we present the average computation times per iteration for executing the global update \eqref{closed_1} (indicated by blue squares), the local update \eqref{closed_2} (depicted by green stars), and the dual update \eqref{ADMM-3} (represented by yellow diamonds). Subsequently, we compute the average total time per iteration (denoted by red circles) by aggregating these three values. It is noteworthy that with an increasing number of CPUs, we observe a decrease in the time required for the local update, while the times for global and dual updates remain relatively constant. This behavior aligns with expectations, as the augmentation of CPUs for parallel computation expedites the local update process. Nevertheless, it is evident that the scalability of CPU-based computation is constrained by the network size.

To provide a comparative perspective, in the subsequent row of Figure \ref{Figure:cpus_gpus}, we present the average computation times when multiple GPUs are employed for parallel computation. As the number of GPUs increases, we observe a slight increase in the times for local updates, while those for global and dual updates remain stable. This phenomenon is primarily attributed to the utilization of MPI for communication within the GPU computing cluster, necessitating data transfer from GPU to CPU. 

For a more comprehensive evaluation of GPU-based computation, excluding the communication overhead, we exclusively employed a single GPU with multiple threads, as reported in the last row of Figure \ref{Figure:cpus_gpus}. This configuration exhibited notably superior performance compared to parallel CPU computation and exhibited scalability with the network size.
Moreover, the increasing number of threads in a GPU benefits the computation time, especially for the IEEE 8500 instance, which has the largest number of subsystems, each of which has a smaller optimization problem in average (see Table \ref{tab:size_distributed}), compared with other IEEE instances. 
Since each thread solves an optimization problem of each subsystem, the computation time taken by each thread for solving the IEEE 8500 instance is expected to be faster, leading to further computational benefit when increasing the number of threads.

\begin{figure}[!h]
\centering        
\includegraphics[width=0.7\linewidth]{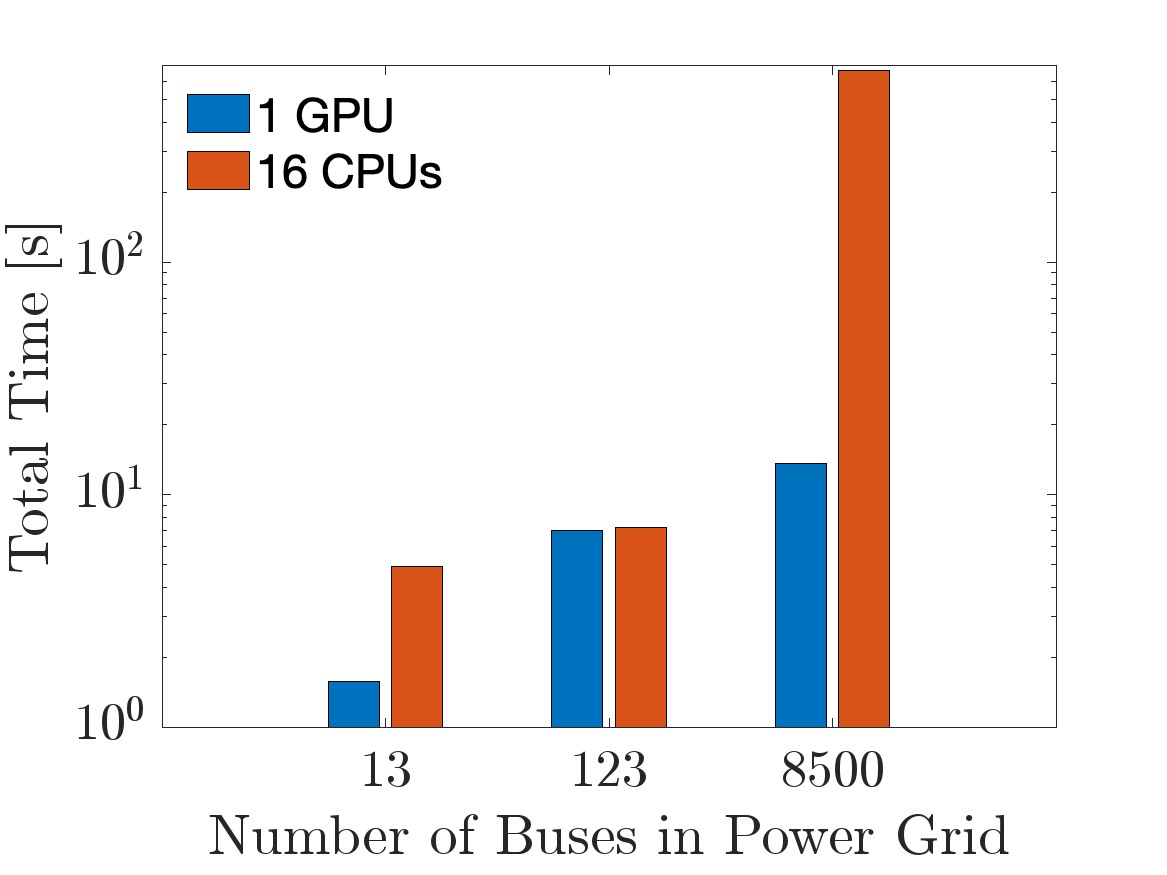}
\caption{Comparison of total time: a GPU versus 16 CPUs. Note that the y-axis is log-scaled.}
\label{fig:time}
\end{figure}

In summary, as presented in Figure \ref{fig:time}, the utilization of a GPU for computation has substantially improved the total computation time, manifesting a fifty-fold enhancement for the IEEE 8500 instance.

\section{Conclusion}
We have introduced a GPU-accelerated distributed optimization algorithm tailored for solving the distributed multi-phase OPF problem within electric power distribution systems. This algorithm builds upon the principles of ADMM and is constructed with subproblems characterized by closed-form expressions relying on matrix operations that can be executed with remarkable efficiency using GPUs.
Notably, the implementation of our approach for GPU has resulted in a substantial reduction in computation time per iteration of the algorithm, when compared to its CPU-based counterparts, particularly evident when applied to large-scale test instances.
Moreover, the speedup achieved by GPU would be significantly increasing with much larger instances.

As part of our future research endeavors, we envision the development of a GPU-accelerated distributed optimization algorithm specifically tailored for the convex relaxation of the multi-phase OPF model. This extension aims to further advance the capabilities of GPU-accelerated optimization techniques in the context of power system analysis and control, potentially opening new avenues for optimizing complex power system problems with enhanced efficiency and scalability.
GPU optimization solvers will also enable the seamless integration of various machine learning techniques (e.g., \cite{zeng2022reinforcement}) that are mainly run on GPUs.

The limitation of the proposed approach is the requirement of the central server aggregating all the local information, potentially raising privacy concern and communication burden. To mitigate such issues, differential privacy \cite{ryu2023differentially} and compression techniques \cite{wilkins2023efficient} could be leveraged in the future.



   
\bibliographystyle{IEEEtran}
\bibliography{reference.bib}

\end{document}